\documentclass[12pt]{article}
\usepackage{epsfig}
\newlength{\imagewidth}
\begin{document}
\title{
	Exact relativistic treatment of stationary 
	counter-rotating dust disks I\\
	Boundary value problems and solutions
}
\author{C.~Klein, 
\\ Laboratoire de Gravitation et Cosmologie 
Relativistes,\\ Universit\'e P.~et M.~Curie, \\
4, place Jussieu, 75005 
Paris, France\\
and \\
Institut f\"ur Theoretische Physik, Universit\"at 
T\"ubingen, \\
Auf der Morgenstelle 14, 72076 T\"ubingen,\\  Germany}
\date{}	

\maketitle

\begin{abstract}
This is the first in a series of papers on the construction of explicit 
solutions to the stationary axisymmetric Einstein equations which 
describe counter-rotating disks of dust.
These disks  can serve as models for certain 
galaxies and accretion disks 
in astrophysics. We review the Newtonian theory for disks using Riemann-Hilbert 
methods which can be extended to some extent to the relativistic 
case where they lead to modular functions on Riemann surfaces. In the 
case of compact surfaces these are  Korotkin's finite gap solutions 
which we will discuss in this paper. On the axis we 
establish  for general genus 
relations between the metric functions and hence the 
multipoles  which are enforced by the underlying hyperelliptic Riemann 
surface. Generalizing these results
to the whole spacetime we are able in principle
to study the classes of boundary value problems which can be solved 
on a given Riemann surface. We investigate the cases of genus 1 and 2 
of the Riemann surface 
in detail and construct the explicit solution for a family of disks 
with constant angular velocity and constant relative energy density 
which was announced in a previous Physical Review Letter.
\end{abstract}

\section{Introduction}\label{sec.1}
The importance of stationary axisymmetric spacetimes arises from the 
fact that they can describe stars and galaxies in thermodynamical 
equilibrium  (see e.g.\ \cite{hartle,lindblom}). However the complicated 
structure of the Einstein equations in the matter region which are 
apparently not completely 
integrable has made a general treatment of these equations 
impossible up to now. Thus only special, possibly unphysical  solutions like 
the one of Wahlquist \cite{wahl} were found (in \cite{perjes} it was 
shown that the Wahlquist solution cannot be the interior solution for 
a slowly rotating star). Since the vacuum 
equations in the form of Ernst \cite{ernst} are known to be 
completely integrable \cite{maison,belzak,neuglinear}, the study of 
two-dimensional matter models can lead to global solutions of the 
Einstein equations which hold both in the matter and in the vacuum 
region: the equations in the matter, which is in general approximated 
as an ideal fluid, reduce to ordinary non-linear differential 
equations because one of the spatial dimensions is suppressed.
The matter thus leads to boundary values for the vacuum equations. 

Disks of pressureless matter, so-called dust, are studied in 
astrophysics as models for certain galaxies and for 
accretion disks. We will therefore discuss 
dust disks in more detail, but the used techniques can in principle 
be extended to more general cases. In the context of galaxy models, 
relativistic effects only play an important role  in the 
presence of black-holes since the latter are genuinely relativistic objects. 
A complete understanding of the black-hole disk system even in 
non-active galaxies is therefore
merely possible in a relativistic setting. The precondition to 
construct exact solutions for stationary black-hole disk systems is 
the ability to treat relativistic disks explicitly.
In this article we will focus on disks of pressureless 
matter. By 
constructing explicit solutions, we hope to get a better understanding of 
the mathematical structure of the field equations and 
the physics of rapidly rotating 
relativistic bodies since dust disks can be viewed as a limiting case for 
extended matter sources.  Hence we will discuss relativistic 
effects for models whose Newtonian limit 
is of astrophysical importance. We will investigate 
disks with counter-rotating dust streams 
which are discussed as models for certain $S0$ and $Sa$ galaxies (see 
\cite{galaxies} and references given therein and \cite{bicak,ledvinka}). 
These galaxies show 
counter-rotating matter components and are believed to be the 
consequence of the merger of galaxies. Recent investigations have 
shown that there is a large number of galaxies (see \cite{galaxies}, 
the first was NGC 4550 in Virgo) which show counter-rotating streams 
in the disk with up to 50 \% 
counter-rotation.

Exact solutions describing relativistic disks are also of interest 
in the context of numerics. They can be used to test existing codes 
for stationary axisymmetric stars as in \cite{eric,lanza}. 
Since Newtonian dust 
disks are known to be unstable against fragmentation and since numerical 
investigations
 (see e.g.\ \cite{bawa}) indicate that the same holds in the relativistic 
case, such solutions could be taken as exact initial data for numerical 
collapse calculations: due to the inevitable numerical error such an 
unstable object will collapse if used as initial data. 

In the Newtonian case, dust disks can be treated in full generality (see e.g.\ 
\cite{binney}) since the disks lead to boundary value problems for the 
Laplace equations which can be solved explicitly. The fact 
that the complex Ernst equation which takes the role of the Laplace equation 
in the relativistic case is completely integrable gives rise to the 
hope that boundary value problems might be solvable here at least in 
selected cases. The unifying framework for both the Laplace and the 
Ernst equation is provided by methods from soliton theory, so-called 
Riemann-Hilbert problems: the scalar problem for the Laplace 
equation can be always solved with the help of a generalization of 
the Cauchy integral (see \cite{jgp2} and references given therein), a procedure 
which leads to the Poisson integral for distributional densities. 
Choosing the contour of the Riemann-Hilbert problem appropriately one can 
construct solutions to the Laplace equation which are everywhere 
regular except at a disk where the function is not differentiable. 
Similarly one can treat the relativistic case where the matrix 
Riemann-Hilbert problem can be related to a linear integral equation.
It was shown in \cite{prd} that 
the matrix problem for the Ernst equation can be always gauge 
transformed to a scalar problem on a Riemann surface which can be 
solved explicitly in terms of 
Korotkin's finite gap solutions \cite{korot1} 
for rational Riemann-Hilbert data. In this sense these 
solutions can be viewed as a generalization of the Poisson integral 
to the relativistic case. 

Whereas the Poisson integral contains one 
free function which is sufficient to solve boundary value problems for 
the scalar gravitational potential, the finite gap solutions contain 
one free function and a set of complex parameters, the branch points of 
the Riemann surface. Thus one cannot hope to solve general boundary 
value problems for the complex Ernst potential within this class
because this would 
imply the choice to specify two free functions in the solution 
according to the boundary data. This means that one can only solve 
certain classes of boundary value problems on a given compact Riemann surface. 
In the first article we investigate the implications of the underlying 
Riemann surface on the multipole moments and the boundary values 
taken at a given boundary. The relations will be given for general 
genus of the surface and will be discussed in detail in the case of genus 
1 (elliptic surface) and genus 2,  which is the 
simplest case with generic equatorial symmetry. It is shown that 
the solution of boundary value problems leads in general to non-linear 
integral equations. We can identify however classes of boundary data 
where only one linear integral equation has to be solved. Special 
attention will be paid to counter-rotating dust disks which will lead us to the 
construction of the solution for constant angular velocity and constant
relative density which was presented in \cite{prl2}. It contains 
as limiting cases the 
static solutions of Morgan and Morgan \cite{morgan} and the disk with 
only one matter stream by Neugebauer and Meinel 
\cite{neugebauermeinel1}.  The potentials of the resulting spacetime 
at the axis and the disk are presented in the second article, the physical 
features as the ultrarelativistic limit, the
formation of ergospheres, multipole moments and the energy-momentum 
tensor are discussed in the third article.

The present article is organized as follows. In section \ref{sec.2}
we discuss Newtonian dust disks with Riemann-Hilbert methods and 
relate the corresponding boundary value problems to an Abelian 
integral equation. The relativistic field equations and the boundary 
conditions for counter-rotating dust disks are summarized in 
section  
\ref{sec.3}. Important facts on hyperelliptic Riemann surfaces which 
will be used to discuss Korotkin's 
class of solutions to the Ernst equation are collected in section \ref{hyper}. 
In section \ref{sec.4}, we establish relations for the corresponding
Ernst potentials on the axis on a given Riemann-surface of arbitrary 
genus. The found 
relation limits the possible choice of the multipole moments. 
We discuss in detail the elliptic and the genus 2 case with equatorial 
symmetry. This analysis is extended to the whole spacetime in section 
\ref{sec.5} which leads to a set of differential and algebraic 
equations which is again discussed in detail for genus 1 and 2. The 
equations for genus 2 are used to study differentially counter-rotating dust 
disks in section \ref{sec.6}: 
We discuss the  Newtonian limit of  disks of genus 2.  
As a first application of this constructive approach we derive 
the class of counter-rotating dust disks with constant angular 
velocity and constant relative density of \cite{prl2}. We prove the 
regularity of the solution up to the ultrarelativistic limit 
in the whole spacetime except the disk and 
conclude in section \ref{conclusion}.

\section{Newtonian dust disks}\label{sec.2}
To illustrate the basic concepts used in the following sections, we 
will briefly recall some facts on Newtonian dust disks. In 
Newtonian theory, gravitation is described by a scalar potential $U$ 
which is a solution to the Laplace equation in the vacuum region. 
We use cylindrical coordinates $\rho$, $\zeta$ 
and $\phi$ and place the disk made up of a pressureless 
two-dimensional ideal fluid with radius $\rho_{0}$ in the equatorial 
plane $\zeta=0$.  In 
Newtonian theory stationary perfect fluid solutions and thus also the 
here considered disks are known to be equatorially symmetric.

Since we concentrate on dust disks, i.e.\ pressureless 
matter, the only force to compensate gravitational attraction 
in the disk is the centrifugal force. This leads in the disk to (here 
and in the following $f_{x}=\frac{\partial f}{\partial x}$)
\begin{equation}
	U_{\rho}=\Omega^{2}(\rho)\rho,
	\label{eq1}
\end{equation}
where $\Omega(\rho)$ is the angular velocity of the dust at radius 
$\rho$. Since all terms in (\ref{eq1}) are quadratic in $\Omega$ 
there are no effects due to the sign of the angular velocity. The 
absence of  
these so-called gravitomagnetic effects in Newtonian 
theory implies that disks with counter-rotating components 
will behave with respect to gravity exactly as 
disks which are made up of only one component. We will therefore only 
consider the case of one component in this section.
Integrating (\ref{eq1}) we get the boundary data $U(\rho,0)$ with an
integration constant $U_{0}=U(0,0)$ which is related to the central redshift in 
the relativistic case. 

To find the Newtonian solution for a given rotation law $\Omega(\rho)$, we 
thus have to construct a solution to the Laplace equation which is 
everywhere regular except at the disk where it has to take on the 
boundary data (\ref{eq1}). At the disk the normal 
derivatives of the potential will have a jump since the disk is a 
surface layer. Notice that one only has to solve the vacuum equations 
since the two-dimensional matter distribution merely leads to boundary 
conditions for the Laplace equation. In the Newtonian setting one thus has 
to determine the density for a given rotation law or vice versa, a 
well known problem (see e.g.\ \cite{binney} and references therein) 
for Newtonian dust disks.

The method we outline here has the advantage that it can be 
generalized to some extent to the relativistic case. We put 
$\rho_{0}=1$ without loss of generality (we are only considering disks 
of finite non-zero radius) and obtain $U$ as the solution of a 
Riemann-Hilbert problem (see e.g.\ \cite{jgp2} and references given 
therein),\\
\textbf{Theorem 2.1:}\\
\emph{Let $\ln G\in C^{1,\alpha}(\Gamma)$  and $\Gamma$ be the covering 
of the 
imaginary axis in the upper sheet of $\Sigma_{0}$ between $-\mathrm{ i}$ 
and $\mathrm{ i}$  where $\Sigma_{0}$ is the Riemann surface of genus 0 given 
by the algebraic relation $\mu_{0}^{2}(\tau)=(\tau-\zeta)^2+\rho^{2}$. 
The function $G$ has to be subject to the conditions 
$G(\bar{\tau})=\bar{G}(\tau)$ and
$G(-\tau)=G(\tau)$. 
Then 
\begin{equation}
	U(\rho,\zeta)=-\frac{1}{4\pi 
	\mathrm{ i}}\int_{\Gamma}^{}\frac{\ln G(\tau) d\tau}{\sqrt{(\tau-\zeta)^2+
	\rho^2}}
	\label{newton2}
\end{equation}
is a real, equatorially symmetric solution to the Laplace equation 
which is everywhere regular except at the disk $\zeta=0$, $\rho\leq 
1$.  The function $\ln G$ is determined by the boundary data 
$U(\rho,0)$ or the energy density $\sigma$ of the dust 
($2\pi \sigma=U_{\zeta}$ in units where the velocity of light and the
Newtonian gravitational constant are equal to 1) via
\begin{equation}
	\ln G(t) = 4\left(U_0+t\int_{0}^{t}
		\frac{U_{\rho}(\rho)d\rho}{\sqrt{t^2-\rho^2}}\right)
	\label{eq4}
\end{equation}
or
\begin{equation}
	\ln G(t)=4 \int_{t}^{1}\frac{\rho U_{\zeta}}{\sqrt{\rho^{2}-t^{2}}}d\rho
	\label{eq5}
\end{equation}
respectively where $t=-\mathrm{i}\tau$.}\\
The occurrence of the logarithm in (\ref{newton2}) is due to the 
Riemann-Hilbert problem with the help of which the solution to the Laplace 
equation was constructed. We briefly outline the \\
\textbf{Proof:}\\
It may be checked by direct calculation that $U$ in (\ref{newton2}) 
is a solution to the Laplace equation except at the disk. The reality condition on $G$ 
leads to a real potential, whereas the symmetry condition with 
respect to the involution $\tau\to -\tau$ leads to equatorial symmetry. 
At the disk the potential takes due to the equatorial symmetry
the boundary values
\begin{equation}
	U(\rho,0)=-\frac{1}{2\pi }\int_{0}^{\rho}
	\frac{\ln G(t) }{\sqrt{\rho^2-t^{2}}}dt
	\label{newton3}
\end{equation}
and 
\begin{equation}
	U_{\zeta}(\rho,0)=-\frac{1}{2\pi 
	}\int_{\rho}^{1}\frac{\partial_{t}(\ln G(t))}{\sqrt{t^{2}-\rho^{2}}}dt
	\label{eq2}.
\end{equation}
Both equations 
constitute integral equations for the `jump data' $\ln G$ of the 
Riemann-Hilbert problem if the respective left-hand side is known. 
The equations (\ref{newton3}) and (\ref{eq2})
are both Abelian integral equations and can be solved in 
terms of quadratures, i.e.\  (\ref{eq4}) and (\ref{eq5}).
To show the regularity of the potential $U$ we prove that the 
integral (\ref{newton2}) is  identical to the 
Poisson integral for a distributional density which reads at the disk
\begin{equation}
	U(\rho)=-2\int_{0}^{1}\sigma(\rho')\rho' d\rho' \int_{0}^{2\pi}
	\frac{d\phi}{\sqrt{(\rho+\rho')^{2}-4\rho\rho' \cos \phi}}
	=-4 \int_{0}^{1}\sigma(\rho')\rho' d\rho'\frac{K(k(\rho,\rho'))}{\rho+\rho'},
	\label{eq7}
\end{equation}
where $k(\rho,\rho')=2\sqrt{\rho\rho'}/(\rho+\rho')$ and where $K$ is the 
complete elliptic integral of the first kind.  Eliminating $\ln G$ in 
(\ref{newton3}) via (\ref{eq5}) we obtain after interchange of the 
order of integration
\begin{equation}
	U=-\frac{2}{\pi}\left(\int_{0}^{\rho}U_{\zeta}\frac{\rho'}{\rho}
	K\left(\frac{\rho'}{\rho}\right)d\rho'+\int_{\rho}^{1}U_{\zeta}
	K\left(\frac{\rho}{\rho'}\right)d\rho'\right)
	\label{eq8}
\end{equation}
which is identical to (\ref{eq7}) since $K(2\sqrt{k}/(1+k))=(1+k)K(k)$. 
Thus the integral (\ref{newton2}) has the 
properties known from the Poisson integral: it is 
a solution to the Laplace equation which is everywhere 
regular except at the disk where the normal derivatives are 
discontinuous. This completes the proof.

\textbf{Remark:} We note that it is possible in the Newtonian case
to solve the boundary value problem purely locally at the disk. 
The regularity properties of the Poisson integral then ensure global 
regularity of the solution except at the disk. Such a purely local 
treatment will not be possible in the relativistic case.

The above considerations make it clear that one cannot prescribe 
both $U$ at the disk (and thus the rotation law) and the density 
independently. This just reflects the fact that the Laplace equation 
is an elliptic equation for which Cauchy problems are ill-posed.
If $\ln G$ is determined by either (\ref{eq4}) or (\ref{eq5}) for 
given rotation law or density, expression 
(\ref{newton2}) gives the analytic continuation of the boundary data 
to the whole spacetime. In case we prescribe the angular velocity, the 
constant $U_{0}$ is determined by the condition $\ln G(\mathrm{i})=0$ which 
excludes a ring singularity at the rim of the disk. For rigid 
rotation ($\Omega=const$), we get e.g.\
\begin{equation}
	\ln G(\tau)=4\Omega^2(\tau^2+1)
	\label{eq6}
\end{equation}
which leads with (\ref{newton2}) to the well-known Maclaurin disk.

\section{Relativistic equations and boundary conditions}\label{sec.3}
It is well known (see \cite{exac}) that the metric of stationary axisymmetric 
vacuum spacetimes can be written in the  Weyl--Lewis--Papapetrou form
\begin{equation}\label{3.1}
	\mathrm{ d} s^2 =-e^{2U}(\mathrm{ d} t+a\mathrm{ d} \phi)^2+e^{-2U}
	\left(e^{2k}(\mathrm{ d} \rho^2+\mathrm{ d} \zeta^2)+
	\rho^2\mathrm{ d} \phi^2\right)
	\label{vac1}
\end{equation}
where $\rho$ and $\zeta$ are Weyl's canonical coordinates and 
$\partial_{t}$ and $\partial_{\phi}$ are the two commuting asymptotically
timelike respectively spacelike Killing vectors. 

In this case the vacuum field equations are equivalent 
to the Ernst equation for the 
complex potential $f$ where $f=e^{2U}+\mathrm{ i}b$, and where
the real function $b$ 
is related to the metric functions via
\begin{equation}\label{3.2}
	b_{z}=-\frac{\mathrm{ i}}{\rho}e^{4U}a_{z}
	\label{vac9}.
\end{equation}
Here the complex variable $z$ stands for $z=\rho+\mathrm{ i}\zeta$. With these
settings, the Ernst equation reads
\begin{equation}\label{3.3}
	f_{z\bar{z}}+\frac{1}{2(z+\bar{z})}(f_{\bar{z}}+f_z)=\frac{2 }{f+\bar{f}}
	f_z f_{\bar{z}}
        \label{vac10}\enspace,
\end{equation}
where a bar denotes complex conjugation in $\mbox{C}$. With a solution $f$,
the metric function $U$ follows directly from the definition of the Ernst 
potential whereas $a$ can be obtained from (\ref{vac9}) via quadratures. 
The metric function $k$ can be calculated from the relation
\begin{equation}\label{3.4}
		k_{z}  =  2\rho \left(U_{z}\right)^2-\frac{1}{2\rho}e^{4U}
                \left(a_{z}\right)^2.
	\label{vac8}
\end{equation}
The integrability condition of (\ref{vac9}) and  (\ref{vac8}) is the 
Ernst equation.
For real $f$, the Ernst equation reduces to the Laplace equation for 
the potential $U$. The corresponding solutions are static and belong 
to the Weyl class. Hence static disks like the counter-rotating disks 
of Morgan and Morgan \cite{morgan} can be treated in the same way as 
the Newtonian disks in the previous section. 

Since the Ernst equation is an elliptic partial differential 
equation, one has to pose boundary value problems. The boundary 
data arise from a solution of the Einstein equations in the 
matter region. In our case this will be an infinitesimally thin disk 
made up of two components of pressureless matter which are counter-rotating.
These models are simple enough that explicit solutions 
can be constructed, and they show typical features of general 
boundary value problems one might consider in the context of the Ernst 
equation. It is also possible to study explicitly the transition from a 
stationary to a static spacetime with a matter source of finite 
extension for these 
models. Counter-rotating disks of infinite extension but finite mass 
were treated in \cite{bicak} and \cite{pichon},  disks producing the 
Kerr metric and other metrics in \cite{ledvinka}.

To obtain the boundary conditions at a relativistic dust disk, it 
seems best to use Israel's invariant 
junction conditions for matching spacetimes across non-null hypersurfaces 
\cite{israel}. Again we place the disk in the equatorial plane and 
match the regions $V^{\pm}$ ($\pm \zeta>0$) at the equatorial plane. 
This is possible with the coordinates of (\ref{vac1}) since we 
are only considering dust i.e.\ vanishing radial
stresses in the disk. The jump $\gamma_{\alpha\beta}=
K^+_{\alpha\beta}-K^-_{\alpha\beta}$ 
in the extrinsic curvature $K_{\alpha\beta}$
of the hypersurface $\zeta=0$ with respect to its embeddings into 
$V^{\pm}=\{\pm\zeta>0\}$ is due to the energy momentum tensor $S_{\alpha\beta}$ of 
the disk via
\begin{equation}
	-8\pi S_{\alpha\beta}=\gamma_{\alpha\beta}-h_{\alpha\beta}
	\gamma_{\epsilon}^{\epsilon}
	\label{vac16.1}
\end{equation}
where $h$ is the metric on the hypersurface (greek indices take the values 
0, 1, 3 corresponding to the coordinates $t$, $\rho$, $\phi$). 
As a consequence of the field 
equations the energy momentum tensor is divergence free, $S^{\alpha\beta}_{;\beta}=0$ 
where the semicolon denotes the covariant derivative with respect to $h$. 

The energy-momentum tensor of the disk is written in the form 
\begin{equation}
	S^{\mu\nu}=\sigma_{+} u^{\mu}_+ u^{\nu}_+ +\sigma_{-}
u^{\mu}_- u^{\nu}_-
	\label{vac16.11},
\end{equation}
where the vectors $u^{\alpha}_{\pm}$ are
a linear combination of the Killing vectors,
$(u^\alpha_{\pm})=(1, 0, \pm\Omega(\rho))$. This 
has to be considered as an algebraic definition of the tensor 
components. Since the vectors $u_{\pm}$ 
are not normalized, the quantities $\sigma_{\pm}$ have no direct 
physical significance, they are just used to parametrize $S^{\mu\nu}$.  
The energy-momentum tensor was chosen in a 
way to interpolate continuously between the static case and the 
one-component case with constant angular velocity. 
An energy-momentum tensor  $S^{\mu\nu}$ with three independent 
components can always be written as 
\begin{equation}
S^{\mu\nu}=\sigma_{p}^{*}v^{\mu}v^{\nu}+p_{p}^{*}w^{\mu}w^{\nu}
\label{2.31a},
\end{equation}
where $v$ and $w$ are the unit timelike respectively spacelike vectors
$(v^{\mu})=N_{1}(1,0,\omega_{\phi})$ and where 
$(w^{\mu})=N_{2}(\kappa,0,1)$.  This corresponds to the introduction 
of observers (called $\phi$-isotropic observers (FIOs) in
\cite{ledvinka}) for which the energy-momentum tensor is diagonal. 
The condition $w_{\mu}v^{\mu}=0$
determines $\kappa$ in terms of $\omega_{\phi}$ and the metric,
\begin{equation}
\kappa=-\frac{g_{03}+\omega_{\phi}g_{33}}{g_{00}+\omega_{\phi}g_{03}}
\label{2.31a1}.
\end{equation}

If $p_{p}^{*}/\sigma_{p}^{*}<1$ the matter in the disk can be 
interpreted as in \cite{morgan} either as having a purely azimuthal 
pressure or as being made up of two counter-rotating streams of 
pressureless matter with proper surface energy density $\sigma_{p}^{*}/2$
which are counter-rotating with the same angular velocity $
\sqrt{p_{p}^{*}/\sigma_{p}^{*}}$,
\begin{equation}
	S^{\mu\nu}=\frac{1}{2}\sigma^{*}(U_{+}^{\mu}U_{+}^{\nu}+
	U_{-}^{\mu}U_{-}^{\nu})
	\label{2.31b}
\end{equation}
where $(U_{\pm}^{\mu})=U^{*}(v^{\mu}\pm \sqrt{p^{*}_{p}/\sigma^{*}_{p}}w^{\mu})$ 
is a unit timelike 
vector. We will always adopt the latter interpretation if the 
condition $p_{p}^{*}/\sigma_{p}^{*}<1$ is satisfied which is the 
case in the example we will discuss in more detail in section 7. 
The energy-momentum tensor (\ref{2.31b}) is just the sum of two 
energy-momentum tensors for dust. Furthermore it can be shown that
the vectors $U_{\pm}$ are geodesic vectors with respect to the inner geometry of 
the disk: this is a consequence of the equation $S^{\mu\nu}_{;\nu}=0$ together 
with the fact that $U_{\pm}$ is a linear combination of the Killing vectors. 
In the discussion of the physical properties of the disk
we will refer only to the measurable quantities 
$\omega_{\phi}$, $\sigma_{p}^{*}$ and $p_{p}^{*}$ which are obtained 
by the introduction of the FIOs whereas $\sigma_{\pm}$ and $\Omega$ 
are just used to generate a sufficiently general energy-momentum 
tensor.

To establish the boundary conditions implied by the energy-momentum 
tensor, we use Israel's formalism \cite{israel}. 
Equation $S^{\alpha\beta}_{;\beta}=0$ leads to the condition 
\begin{equation}
	U_{\rho}\left(1+2\gamma\Omega a+\Omega^{2}a^{2}\right) +
	\Omega a_{\rho} (\gamma+\Omega a) 
	+\Omega^2 \rho (\rho U_{\rho}-1) e^{-4U}=0,
	\label{vac20}
\end{equation}
where 
\begin{equation}
	\gamma(\rho)=\frac{\sigma_{+}(\rho)-\sigma_{-}(\rho)}{\sigma_{+}(\rho)
	+\sigma_{-}(\rho)}
	\label{vac20a}.
\end{equation}
The function $\gamma(\rho)$ is a measure for the relative energy density of the 
counter-rotating matter streams. 
For $\gamma\equiv 1$, there is only one component of matter, for 
$\gamma\equiv 0$, the matter streams have identical density which 
leads to a static spacetime of the Morgan and Morgan class. 

As in the Newtonian case, one cannot prescribe  both the proper energy 
densities 
$\sigma_{\pm}$ and the rotation law $\Omega$ at the disk since the 
Ernst equation is an elliptic equation. For the matter model 
(\ref{vac16.11}), we get at the disk \\
\textbf{Theorem 3.1:}\\
\emph{Let $\tilde{\sigma}(\rho)=\sigma_{+}(\rho)+\sigma_{-}(\rho)$ and let
$R(\rho)$ and $\delta(\rho)$ be given by
\begin{equation}
	R=\left(a+\frac{\gamma}{\Omega}\right)e^{2U}
	\label{eq10},
\end{equation}
and 
\begin{equation}
	\delta(\rho)=\frac{1-\gamma^{2}(\rho)}{\Omega^{2}(\rho)}
	\label{eq9a}.
\end{equation}
Then for prescribed $\Omega(\rho)$ and $\delta(\rho)$, the boundary data at the 
disk take the form 
\begin{equation}
	f_{\zeta}=-\mathrm{i}\frac{R^{2}+\rho^{2}+\delta e^{4U}}{2R\rho}f_{\rho}
	+\frac{\mathrm{i}}{R}e^{2U}
	\label{eq9}.
\end{equation}
Let $\sigma$ be given by
$\sigma=\tilde{\sigma }
e^{k-U}$.
Then for given density $\sigma$ and $\gamma$, the boundary data 
read,
\begin{equation}
	(\rho^2+\delta e^{4U}) \left(\left(e^{2U}\right)_{\rho}
	\left(e^{2U}\right)_{\zeta} +b_{\rho}b_{\zeta}\right)^2
	-2\rho e^{2U} \left(e^{2U}\right)_{\zeta}\left(\left(e^{2U}\right)_{\rho}
	\left(e^{2U}\right)_{\zeta} +b_{\rho}b_{\zeta}\right)
	+b_{\rho}^2 e^{4U}=0
	\label{16.15},
\end{equation}
and
\begin{equation}
	\left(b_{\rho}-a\left(\left(e^{2U}\right)_{\rho}
	\left(e^{2U}\right)_{\zeta} +b_{\rho}b_{\zeta}\right)\right)^2+8\pi 
	\rho\sigma e^{2U}\gamma^{2}\left(\left(e^{2U}\right)_{\rho}
	\left(e^{2U}\right)_{\zeta} +b_{\rho}b_{\zeta}\right)=0
	\label{16.17}.
\end{equation}
}

\textbf{Proof:}\\
The relations (\ref{vac16.1}) lead to 
\begin{eqnarray}
	-4\pi e^{(k-U)}S_{00} & = & \left(k_{\zeta}-2U_{\zeta}
	\right)e^{2U},
	\nonumber \\
    -4\pi e^{(k-U)}( S_{03}-aS_{00}) & = & -\frac{1}{2}a_{\zeta}e^{2U}
	\nonumber ,\\
	-4\pi e^{(k-U)} (S_{33}-2a S_{03}+a^2 S_{00})& = & -k_{\zeta}\rho^2e^{-2U}
	\label{16.7},
\end{eqnarray}
where 
\begin{eqnarray}
	S_{00} & = & \tilde{\sigma} e^{4U}\left(1+\Omega^2 a^2 +2\Omega a \gamma
	\right),
	\nonumber \\
	S_{03}-a S_{00} & = & -\tilde{\sigma} \rho^2 \Omega \left(\Omega a +\gamma
	\right),
	\nonumber \\
	S_{33}-2a S_{03}+a^2 S_{00} & = & \tilde{\sigma} \Omega^2 \rho^4 e^{-4U}
	\label{gegen3}.
\end{eqnarray}
One can substitute one of the above equations by (\ref{vac20}) in the 
same way as one replaces one of the field equations by the covariant 
conservation of the energy momentum tensor in the case of 
three-dimensional ideal fluids. This makes it possible to eliminate 
$k_{\zeta}$ from (\ref{16.7}) and to treat the 
boundary value problem purely on the level of the Ernst equation. The 
function $k$ will then be determined
via (\ref{vac8}) with the found solution of the Ernst equation. It is 
straight forward to check 
the consistency of this approach with the help of (\ref{vac8}).

If  $\Omega$ and $\gamma$ (and thus $\delta$) are given, one has to eliminate 
$\tilde{\sigma}$ from (\ref{16.7}) and (\ref{gegen3}).
This can be combined with (\ref{vac20}) and (\ref{vac9}) to (\ref{eq9}).

If the function $\gamma$ and $\sigma$  are prescribed (this 
makes it possible to treat the problem completely on the level of the 
Ernst equation), one has to eliminate $\Omega$ from (\ref{vac20}), 
(\ref{16.7}) and (\ref{gegen3}) which leads to (\ref{16.15}) and 
(\ref{16.17}). This completes the proof.

\textbf{Remark:}
For given $\Omega(\rho)$ and $\delta(\rho)$, 
equation (\ref{vac20}) is an ordinary non-linear 
differential equation for $e^{2U}$,
\begin{equation}
	(R^{2}-\rho^{2})_{\rho}e^{2U}-2R e^{4U}
	\left(\frac{\gamma}{\Omega}\right)_{\rho} =(R^{2}-\rho^{2}-\delta 
	e^{4U})\left(e^{2U}\right)_{\rho}
	\label{vac8a}.
\end{equation}
For constant $\Omega$ and $\gamma$
we get 
\begin{equation}
	R^{2}-\rho^{2}+\delta e^{4U}=\frac{2}{\lambda}e^{2U}
	\label{eq11},
\end{equation}
where $\lambda=2\Omega^{2}e^{-2U_{0}}$.

For given boundary  values as in Theorem 
3.1, the task is to  to find a solution to the Ernst equation which 
is regular in the whole spacetime except at the disk where it 
has to satisfy two real boundary conditions. In the following we will 
concentrate on the case where the angular velocity $\Omega$ and the relative 
density $\gamma$ are prescribed.

\section{Solutions on hyperelliptic Riemann surfaces}\label{hyper}
The remarkable feature of the Ernst equation is that it is completely 
integrable which means that the Riemann-Hilbert techniques used in the 
Newtonian case can be applied here, too. This time, however, one has 
to solve a matrix problem (see e.g.\ \cite{prd} and references given 
therein) which cannot be solved generally in closed form. In 
\cite{prd} it was shown that the problem can be gauge transformed to a 
scalar problem on a four-sheeted 
Riemann surface. In the case of rational `jump data' 
of the Riemann-Hilbert problem, this surface is compact and the 
corresponding solutions to the Ernst equation are Korotkin's finite 
gap solutions \cite{korot1}. In the following we will concentrate on 
this class of solutions and investigate its properties with respect to 
the solution of boundary value problems.

\subsection{Theta functions on hyperelliptic Riemann surfaces}\label{subsec.1}
We will first summarize some basic facts on hyperelliptic Riemann 
surfaces which we will need in the following. We consider
surfaces $\Sigma$ of genus $g$ which are
given by the relation $\mu^{2}(K)=(K+\mathrm{ i}z)(K-\mathrm{ i}\bar{z})
\prod_{i=1}^{g}(K-E_{i})(K-\bar{E}_{i})$ where the $E_{i}$ do not 
depend on the physical coordinates $z$ and $\bar{z}$. We 
introduce the standard quantities associated with a Riemann surface 
(see \cite{farkas}), with respect to the cut system of figure 1 (we 
order the branch points with $\mbox{Im}E_{i}<0$ in a way that 
$\mbox{Re}E_{1}<\mbox{Re}E_{2}<\ldots <\mbox{Re}E_{g}$ and assume for 
simplicity that the real parts of the $E_{i}$ are all different; we 
write $E_{i}=\alpha_{i}+\beta_{i}$), 
the $g$ normalized differentials of
the first kind $\mathrm{ d}\omega_i$ defined by $\oint_{a_i}\mathrm{ d}\omega_j=
2\pi\mathrm{ i} 
\delta_{ij}$, and with $P_{0}=-\mathrm{i}z$
the Abel map $\omega_i(P)=\int_{P_0}^{P}\mathrm{ d}\omega_i$ which is 
defined uniquely up to periods.
Furthermore, we define
the Riemann matrix $\Pi$ with the elements $\pi_{ij}=
\oint_{b_i}\mathrm{ d}\omega_j$, and the theta function 
$\Theta\left[m\right](z)=
\sum_{N\in{Z}^g}^{}\exp\left\{\frac{1}{2}\left\langle\Pi (N+
\frac{m^{1}}{2}),(N+\frac{m^{1}}{2})
\right\rangle+\left\langle
(z+\pi \mathrm{ i} m^{2}),(N+\frac{\alpha}{2})
\right\rangle\right\}$ 
with half integer
characteristic $[m]=\left[m^{1}
\atop m^{2}\right]$ and $m^{1}_i,m^{2}_i=0,1$ 
($\left\langle N,z\right\rangle=\sum_{i=1}^g N_iz_i$). 
A characteristic is called 
odd if $\langle m^{1},m^{2}\rangle \neq 0 \mbox{ mod } 2$. The 
normalized (all $a$--periods zero) differential of the third kind with 
poles at $P_{1}$ and $P_{2}$  and residue $+1$ and 
$-1$  respectively will be denoted by $\mathrm{ d} 
\omega_{P_{1}P_{2}}$.  A point $P\in \Sigma$ will be denoted by 
$P=(K,\pm \mu(K))$ or $K^{\pm}$ (the sheets will be defined in the 
vicinity of a given point on $\Sigma$, e.g.\ $\infty$). 

\begin{figure}[tbp]
	\centering
	\epsfig{file=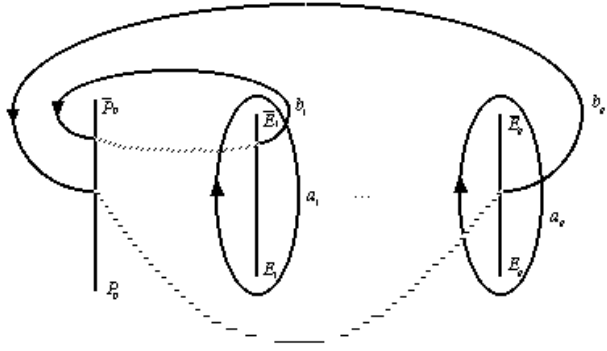,width=\imagewidth}\\[7pt] 
	\caption{Homology basis of $\Sigma$}
	\label{fig1}
\end{figure}

The theta functions are subject to a number of addition theorems. We 
will need the ternary addition theorem which can be cast in the form\\
\textbf{Theorem 4.1:} Ternary addition theorem\\
\emph{Let $[m_{i}]=[m_{i}^{1},m_{i}^{2}]$ $(i=1,\ldots,4)$ be 
arbitrary real $2g$-dimensional vectors. Then
\begin{eqnarray}
	&&\Theta[m_1](u+v)\Theta[m_2](u-v)\Theta[m_3](0)\Theta[m_4](0)
	\label{ternary}\\
	&&=\frac{1}{2^g}\sum\limits_{2a \in (Z_2)^{2g}}\exp(-4\pi \mathrm{i} \langle
	m_1^{1},a^{2}\rangle )
	\Theta[n_1+a](u)\Theta[n_2+a](u)\Theta[n_3+a](v)\Theta[n_4+a](v)
	\nonumber,
\end{eqnarray}
where $a=(a^{1},a^{2})$,  and
$(m_1,\dots,m_4)=(n_1,\dots,n_4) T$
with
\begin{equation}
	T=\frac{1}{2}\left (
	\begin{array}{rrrr}
		1 & 1 & 1 & 1  \\
		1 & -1 & -1 & -1  \\
		1 & -1 & 1 & -1  \\
		1 & -1 & -1 & 1
	\end{array}
	\right)
\label{T}.
\end{equation}
Each $1$ in $T$ denotes the $g\times g$ identity matrix.}

For a proof see e.g.\ \cite{algebro}. 

Let us recall that a divisor $X$ on $\Sigma$
is a formal symbol  $X=n_{1}P_{1}+\ldots+ n_{k}P_{k}$ with $P_{i}\in 
\Sigma$ and $n_{i}\in \mbox{Z}$. The degree of a divisor is 
$\sum_{i=1}^{k}n_{i}$.  The Riemann vector $K_{R}$ is 
defined by the condition that $\Theta(\omega(W)+K_{R})=0$ if $W$ is a 
divisor of degree $g-1$ or less. We use here and in the following the 
notation $\omega(W)=\int_{P_{0}}^{W}\mathrm{d}\omega
=\sum_{i=1}^{g-1}\omega(w_{i})$. 
We note that  the Riemann vector can be 
expressed through half-periods in the case of a hyperelliptic surface.

The quotient of two theta functions with the same argument but 
different characteristic is a so-called root function which means that 
its square is a function on $\Sigma$. One can prove (see 
\cite{algebro} and references therein) \\
\textbf{Theorem 4.2:} Root functions\\
\emph{Let  $Q_{i}$, $i=1,\ldots, 2g+2$, be 
the branch points of a hyperelliptic Riemann surface $\Sigma_{g}$ of genus $g$ 
and $A_{j}=\omega(Q_{j})$ with $\omega(Q_{1})=0$. Furthermore let 
$\{i_{1},\ldots, i_{g}\}$ and $\{j_{1},\ldots,j_{g}\}$ be two sets of 
numbers in $\{1,2,\ldots,2g+2\}$. Then the following 
equality holds for an 
arbitrary point $P\in \Sigma_{g}$,
\begin{equation}
	\frac{\Theta\left[K_{R}+\sum_{k=1}^{g}A_{i_{k}}\right]\left(
	\omega(P)\right)}{\Theta\left[K_{R}+
	\sum_{k=1}^{g}A_{j_{k}}\right]\left(
	\omega(P)\right)}=c_{1}\sqrt{\frac{(K-E_{i_{1}})\ldots(K-E_{i_{g}})}{
	(K-E_{j_{1}})\ldots(K-E_{j_{g}})}}
	\label{root1},
\end{equation}
where $c_{1}$ is a constant independent on $K$. Let 
$X=P_{1}+\ldots+P_{g}$ with $P_{j}=(K_{j},\mu(K_{j})$ be a divisor of 
degree g on $\Sigma_{g}$ then the following identity exists,
\begin{equation}
	\frac{\Theta\left[K_{R}+A_{i}\right]\left(
	\omega(X)\right)}{\Theta\left[K_{R}+
	A_{j}\right]\left(
	\omega(X)\right)}=c_{2}\prod_{k=1}^{g}
	\sqrt{\frac{(K_{k}-Q_{i})}{(K_{k}-Q_{j})
	}}
	\label{root2},
\end{equation}
where $c_{2}$ is a constant independent on the $K_{k}$.
}

The notion of divisors makes it possible to state Jacobi's inversion 
theorem in a very compact form,\\
\textbf{Theorem 4.3:}
	Jacobi inversion theorem \\
	\emph{ Let $A,B \in \Sigma$ be divisors of degree g and $u\in 
	\mathrm{C}^{g}$. Then for given $B$ and $u$, the equation 
	$\omega(A)-\omega(B)=u$ for the divisor $A$ is always solvable.}

For a proof we refer the reader to the standard literature, e.g.\ 
\cite{farkas}. We remark that the divisor may not be uniquely defined 
in the general case which means that one or more $P_{i}\in A$ can be 
freely chosen. We will not consider such special cases in the 
following and refer the reader for the so-called special divisors 
to the literature as \cite{algebro}. 

For divisors $A-B$ of degree zero, one 
can formulate Abel's theorem.\\
\textbf{ Theorem 4.4:} Abel's theorem \\
\emph{ Let $A,B\in \Sigma$ be divisors of degree $n$ subject to the 
relation $\omega(A)-\omega(B)=0$. Then $A$ and $B$ are the set of 
zeros respectively poles of a meromorphic function $F$.}

For a proof see \cite{farkas}. We remark that this function is a 
rational function on the surface cut along the homology basis. We have the 

\textbf{ Corollary 4.5:}\\
\emph{ Let the condition of Abel's theorem hold. Then the following 
identity holds for the 
integral of the third kind
\begin{equation}
	\int_{B}^{A}d\omega_{PQ}=\ln \frac{F(P)}{F(Q)}
	\label{eq20}.
\end{equation}
}

\subsection{Solutions to the Ernst equation}\label{subsec.4.2}
We are now able to write down a class of solutions to the Ernst 
equation on the surface $\Sigma$.\\
\textbf{ Theorem 4.5:} \\
\emph{ Let the Riemann surface $\Sigma$ be given by the relation 
$\mu^{2}(K)=(K+\mathrm{i}z)(K-\mathrm{i}z)\prod_{i=1}^{g}(K-E_{i})
(K-\bar{E}_{i})$, let
$u$ be the vector with the components $u_i=\frac{1}{2\pi \mathrm{ i}}
\int_{\Gamma}^{}\ln G d\omega_i$ where $\Gamma$ is as in theorem 2.1, 
let $G$ be subject to the condition $G(\tau)=\bar{G}(\bar{\tau})$, 
and let  $[m]=[m^{1},m^{2}]$ with 
$m^{1}_{i}=0$ and $m^{2}_{i}$ arbitrary for $i=1,\ldots,g$ be a theta 
characteristic. 
Then the function $f$ given by
\begin{equation}
	f(\rho,\zeta)=\frac{\Theta[m](\omega(\infty^{+})+u)}{
	\Theta[m](\omega(\infty^{-})+u)}
\exp\left\{
\frac{1}{2\pi\mathrm{ i}}\int\limits_\Gamma
\ln G(\tau)\mathrm{ d}\omega_{\infty^{+}\infty^-}(\tau)
\right\}
\label{rel1a},
\end{equation}
is a solution to the Ernst equation.}\\ 
This class of solutions was first given by Korotkin \cite{korot1}, the 
straight forward
continuous limit leading to the above form can be found in 
\cite{korotmat1,korotmat2}. 
For the relation to Riemann-Hilbert 
problems see \cite{prd}. In the case genus 0, the Ernst 
potential is real, and we get a solution of the Weyl class in the form 
(\ref{newton2}). For higher genus, these solutions are in general 
non-static and thus generalize (\ref{newton2}) to the stationary 
case. 

In \cite{prl,prd2} it was possible to identify a physically 
interesting subclass.

\textbf{Theorem 4.6:}\\
\emph{Let the conditions of Theorem 4.5 hold, and in addition let 
$\Sigma$ be a hyperelliptic Riemann surface of even genus 
$g=2n$
given by 
$\mu^{2}(K)=(K+\mathrm{i}z)(K-\mathrm{i}z)\prod_{i=1}^{n}(K^{2}-E_{i}^{2})
(K^{2}-\bar{E}_{i}^{2})$, let 
the function $G$ be subject to the 
condition $G(-\tau)=G(\tau)$, 
and let $[n]$ be the  
characteristic with $n^{1}_{i}=0$ and $n^{2}_{i}=1$. 
Then the function $f$ given by
\begin{equation}
	f(\rho,\zeta)=\frac{\Theta[n](\omega(\infty^{+})+u)}{
	\Theta[n](\omega(\infty^{-})+u)}
\exp\left\{
\frac{1}{2\pi\mathrm{ i}}\int\limits_\Gamma
\ln G(\tau)\mathrm{ d}\omega_{\infty^{+}\infty^-}(\tau)
\right\}
\label{rel1}
\end{equation}
is an equatorially symmetric solution to the Ernst equation 
($f(-\zeta)=\bar{f}(\zeta)$) which is everywhere regular except at 
 the disk if $\Theta(\omega(\infty^{-})+u) \neq 0$.}
 
For a proof see \cite{prl,prd2} where one can also find how the 
characteristic can be generalized. In the following we will only use 
the characteristic of the above theorem.

A quantity of special interest is the metric function $a$. In 
\cite{korot1} it was shown that one can relate it directly to 
theta functions without having to perform the integration of 
(\ref{vac9}),
\begin{equation}
	Z:=(a-a_{0})e^{2U}=D_{\infty^{-}}\ln\frac{\Theta(\omega(\infty^{-})+u)}{
	\Theta[n](\omega(\infty^{-})+u)}
	\label{eq24}
\end{equation}
where $D_{P}F(\omega(P))$ denotes the coefficient of the linear 
term in the expansion of 
the function $F(\omega(P))$ in the local parameter in the 
neighborhood of $P$, where $\Theta$ is the Riemann theta function 
with the characteristic $[m]$ and $m^{1}_{i}=m^{2}_{i}=0$, 
and where the constant $a_{0}$ is defined by 
the condition that $a$ vanishes on the regular part of the axis.

It is possible to give an algebraic 
representation of the solutions (\ref{rel1}) (see \cite{meinelneugebauer} and 
\cite{korotneu}). 
We define the divisor $X=\sum_{i=1}^{g}K_{i}$ 
as the solution of the Jacobi inversion problem ($i=1,\ldots,g$)
\begin{equation}
	\omega_{i}(X)-\omega_{i}(D)=\frac{1}{2\pi \mathrm{i}}
	\int_{\Gamma}^{}\ln G \frac{\tau^{i-1}d\tau}{\mu(\tau)}=:\tilde{u}_{i}
	\label{eq18},
\end{equation}
where the divisor $D=\sum_{i=1}^{g}E_{i}$. 
With the help of these divisors, we can write (\ref{rel1}) in the form
\begin{equation}
	\ln f=\int_{D}^{X}\frac{\tau^{g}d\tau}{\mu(\tau)}-\frac{1}{2\pi \mathrm{i}}
	\int_{\Gamma}^{}\ln G \frac{\tau^{g}d\tau}{\mu(\tau)}
	\label{eq19},
\end{equation}

Since the $\tilde{u}_{i}$ in (\ref{eq18})
are just the periods of the second integral in 
(\ref{eq19}), they are subject to a system of differential equations, 
the so-called Picard-Fuchs system (see \cite{prd2} and references 
given therein). In our case this leads to  
\begin{equation}
	\sum_{n=1}^{g}\frac{(K_n-P_0)K_n^j}{\mu(K_n)}K_{n,z}=0, \quad j=0,..., g-2
	\label{ddd17}
\end{equation}
and 
\begin{equation}
	(\ln f)_{z}=\sum_{n=1}^{g}\frac{(K_n-P_0)K_n^{g-1}}{\mu(K_n)}K_{n,z}
	\label{ddd18}.
\end{equation}

Solving for the $K_{n,z}$, $n=1,\ldots,g$, we get 
\begin{equation}
	K_{n,z}=(\ln 
	f)_{z}\frac{\mu(K_{n})}{K_{n}-P_{0}}\frac{1}{\prod_{m=1, m\neq 
	n}^{g}(K_{n}-K_{m})}
	\label{ddd19}.
\end{equation}

Additional information follows from the reality of the $\tilde{u}_{i}$ 
which implies $\omega(X)-\omega(D)=\omega(\bar{X})-\omega(\bar{D})$. 
Using Abel's theorem on this condition, 
we obtain the relation for an arbitrary $K\in \mbox{C}$
\begin{equation}
	(1-x^2)\prod_{i=1}^{g}(K-K_i)(K-\bar{K}_i)=\prod_{i=1}^{g}(K-E_i)(K 
	-\bar{E}_i)-(K-P_0)(K-\bar{P}_0)Q_2^2(K)
	\label{ddd2},
\end{equation}
where with purely imaginary $x_i$, $x$
\begin{equation}
	Q_2(K)=x_0+x_1K+...+x_{g-2}K^{g-2}+xK^{g-1}
	\label{ddd3}.
\end{equation}
Since (\ref{ddd2}) has to hold for all $K\in\mbox{C}$, it is equivalent 
to $2g$ real  algebraic equations 
for the $K_{i}$ if the $x_{i}$ are given. Using (\ref{eq20}) and 
(\ref{eq19}) we find
\begin{equation}
	\frac{f}{\bar{f}}=\frac{1+x}{1-x}
	\label{eq21}
\end{equation}
which implies $x=\mathrm{i}be^{-2U}$.

\textbf{Remark:} To solve boundary value problems with the class of solutions 
(\ref{rel1}), one has two kinds of freedom: 
the function $G$ as before and the branch points 
$E_{i}$ of the Riemann surface as a discrete degree of freedom. Since 
one would need to specify two free functions to solve a 
general boundary value problem for the Ernst equation, it is obvious 
that one can only solve a restricted class of problems on a given surface, and 
that one cannot expect to solve general problems on a surface of 
finite genus. But once one has constructed a 
solution which takes the imposed boundary data at the disk, one has 
to check the condition $\Theta(\omega(\infty^{-})+u) \neq 0$ in the 
whole spacetime to actually prove that one has found the 
desired solution: a solution that is everywhere regular except at 
the disk where it has to take the imposed boundary conditions.

There are in principle two ways of generalizing the 
approach used for the Newtonian case: One can eliminate $\Omega$ from 
the two real equations (\ref{eq9}) and enter the resulting equation
with a solution 
(\ref{rel1}) on a chosen Riemann surface. This will lead for given 
$\gamma$ to a 
non-linear integral equation for $\ln G$. In general there is little 
hope to get explicit solutions to this equation (for a numerical 
treatment of differentially rotating disks along this line 
in the genus 2 case see 
\cite{ansorgmeinel}). Once a function $G$ is found, one can read off 
the rotation law $\Omega$ on a given Riemann surface from (\ref{rel1}).
Another approach is to establish the relations between the real and 
the imaginary part of the Ernst potential which exist on a given 
Riemann surface for arbitrary $G$. The simplest example for such a 
relation is provided by the function $w=e^{\mathrm{i}\psi}$ which is a 
function on a Riemann surface of genus 0, where we have obviously 
$|w|=1$. As we will point out in the following, similar relations 
also exist for an Ernst potential of the form (\ref{rel1}), but they 
will  lead to a system of differential 
equations. Once one has established these 
relations for a given Riemann surface, one can determine in 
principle which boundary value 
problems can be solved there (in our example which classes of 
functions $\Omega$, $\gamma$ can occur) by the condition that one of the boundary 
conditions must be identically satisfied. The second equation will then be 
used to determine $G$ as the solution of an integral equation which is 
possibly non-linear. Following the second approach, we want 
to study the implications of the hyperelliptic Riemann surface for 
the physical properties of the solutions.

\section{Axis Relations}\label{sec.4}
In order to establish relations between the real and the imaginary 
part of the Ernst potential, we will first consider the axis of 
symmetry ($\rho=0$) where the situation simplifies decisively. In 
addition the 
axis is of interest since the asymptotically defined multipole moments 
\cite{geroch,hansen} can be read off there \cite{fodor}.

On the axis the Ernst potential can be expressed through functions 
defined on the Riemann surface $\Sigma'$ given by 
$\mu'{}^{2}=\prod_{i=1}^{g}(K-E_{i})(K-\bar{E}_{i})$, i.e.\ the 
Riemann surface obtained from $\Sigma$ by removing the cut 
$[P_{0},\bar{P}_{0}]$ which just collapses on the axis. We 
use the notation of the previous section and let a prime denote that 
the corresponding quantity is defined on the surface $\Sigma'$. The 
cut system is as in the previous section with $[E_{1}, \bar{E}_{1}]$ 
taking the role of $[P_{0},\bar{P}_{0}]$ (all $b$-cuts cross 
$[E_{1},\bar{E}_{1}]$). We choose the Abel map in a 
way that $\omega'(E_{1})=0$.
It was shown in \cite{prd2} that for genus 
$g>1$ the Ernst potential takes the form (for $\zeta>0$)
\begin{equation}
        f(0,\zeta)=\frac{\vartheta
        \left(\int_{\zeta^+}^{\infty^+}\mathrm{d}\omega'+u'\right)
	-\exp(-(\omega'_g(\infty^+)+u_g))
	\vartheta\left(\int_{\zeta^-}^{\infty^+}\mathrm{d}\omega'
	+u'\right)}{\vartheta
        \left(\int_{\zeta^+}^{\infty^+}\mathrm{d}\omega'-u'\right)
	-\exp(-(\omega'_g(\infty^+)-u_g))
	\vartheta\left(\int_{\zeta^-}^{\infty^+}\mathrm{d}\omega'
	-u'\right)}e^{I+u_{g}}
        \label{sing7},
\end{equation}
where $\vartheta$ is the theta function on $\Sigma'$ with the 
characteristic $\alpha_{i}'=0$, $\beta_{i}'=1$ for $i=1,\ldots, g-1$, 
where $I=\frac{1}{2\pi \mathrm{i}}
        \int\limits_\Gamma
\ln G(\tau)\mathrm{ d}\omega'_{\infty^{+}\infty^-}(\tau)$,
$\mathrm{d}\omega_{g}= \mathrm{d}\omega_{\zeta^{-}\zeta^{+}}$, and where 
$u_{g}=\frac{1}{2\pi \mathrm{i}}
        \int\limits_\Gamma
\ln G(\tau)\mathrm{ d}\omega_{g}(\tau)$.

Notice that the $u_{i}'$ and $I$ are constant with respect to $\zeta$. The 
only term dependent both on $G$ and on $\zeta$ is $u_{g}$. To 
establish a relation on the axis between the real and the imaginary 
part of the Ernst potential independent of $G$, the first step must be 
thus to eliminate $u_{g}$. We can state the following

\textbf{ Theorem 4.1:}\\
\emph{ The Ernst potential (\ref{sing7}) satisfies for $g> 1$
the relation 
\begin{equation}
	P_{1}(\zeta)f\bar{f}+P_{2}(\zeta)b+P_{3}(\zeta)=0
	\label{eq13},
\end{equation}
where the $P_{i}$ are real polynomials in $\zeta$
with coefficients depending on the branch points $E_{i}$ and the 
$g$ real constants $\int_{\Gamma}^{}\ln G \tau^{i}d\tau /\mu'(\tau)$ 
with $i=0,\ldots,g-1$. The degree of the polynomials $P_{1}$ and 
$P_{3}$ is $2g-3$ or less, the degree of $P_{2}$ is $2g-2$ or less.}

To prove this theorem we need the fact
that one can express integrals of the third kind via 
theta functions with odd characteristic denoted by $\vartheta_{o}$,
\begin{equation}
\exp(-\omega_{g}(\infty^{+}))=-\frac{\vartheta_{o}(\omega'(\infty^{+})-
\omega'(\zeta^{+}))}{\vartheta_{o}(\omega'(\infty^{+})-
\omega'(\zeta^{-}))}.
	\label{eq13.3}
\end{equation}
\textbf{ Proof:}\\
The first step is  to establish the relation 
\begin{equation}
	Af\bar{f}+B\mathrm{i}b+1=0
	\label{eq12},
\end{equation}
where 
\begin{equation}
	Ae^{2I}=-\frac{\vartheta\left(u'+\int_{\zeta^{-}}^{\infty^{-}}\mathrm{d}\omega'\right)
	\vartheta\left(u'+\int_{\zeta^{+}}^{\infty^{-}}\mathrm{d}\omega'\right)}{
	\vartheta\left(u'+\int_{\zeta^{-}}^{\infty^{+}}\mathrm{d}\omega'\right)
	\vartheta\left(u'+\int_{\zeta^{+}}^{\infty^{+}}\mathrm{d}\omega'\right)}
	\label{eq14}
\end{equation}
and 
\begin{equation}
	Be^{I}=\frac{e^{-\omega_{g}(\infty^{+})}\vartheta\left(
	u'+\int_{\zeta^{-}}^{\infty^{+}}\mathrm{d}\omega'\right)
	\vartheta\left(u'+\int_{\zeta^{+}}^{\infty^{-}}\mathrm{d}\omega'\right)
	+e^{\omega_{g}(\infty^{+})}\vartheta\left(u'+
	\int_{\zeta^{+}}^{\infty^{+}}\mathrm{d}\omega'\right)
	\vartheta\left(u'+\int_{\zeta^{-}}^{\infty^{-}}\mathrm{d}\omega'\right)}{
	\vartheta\left(u'+\int_{\zeta^{-}}^{\infty^{+}}\mathrm{d}\omega'\right)
	\vartheta\left(u'+\int_{\zeta^{+}}^{\infty^{+}}\mathrm{d}\omega'\right)}
	\label{eq15}
\end{equation}
which may be checked with (\ref{sing7}) by direct calculation. The 
reality properties of the Riemann surface $\Sigma'$ and the function $G$ 
imply that $A$ is real and that $B$ is purely imaginary. We 
use the addition theorem (\ref{ternary}) with $[m_{1}]=\ldots=[m_{4}]$ 
equal to the  characteristic of $\vartheta$ for 
(\ref{eq14}) to get 
\begin{equation}
	Ae^{2I}=-\frac{\sum\limits_{2a \in (Z_2)^{2g}}\exp(-4\pi \mathrm{i} \langle
	m_1^{1},a^{2}\rangle )\vartheta^{2}[a](u'+\omega'(\infty^{-}))
	\vartheta^{2}[a](\omega'(\zeta^{+}))}{\sum\limits_{2a \in 
	(Z_2)^{2g}}\exp(-4\pi \mathrm{i} \langle
	m_1^{1},a^{2}\rangle )
	\vartheta^{2}[a](u'+\omega'(\infty^{+}))
	\vartheta^{2}[a](\omega'(\zeta^{+}))}.
	\label{eq16}
\end{equation}
This term is already in the desired form. 
Using the 
relation for root functions (\ref{root1}), one can directly see that 
the right-hand side is a quotient of polynomials of order $g-1$ or lower 
in $\zeta$. For (\ref{eq15}) we use (\ref{eq13.3}) with 
$[\tilde{m}_{1}]=[\tilde{m}_{2}]
=[K_{R}]$ as 
the characteristic of the odd theta function $\vartheta_{o}$ and let 
$[\tilde{m}_{3}]=[\tilde{m}_{4}]$ be equal to the characteristic of 
$\vartheta$. 
The addition theorem (\ref{ternary})
then leads to 
\begin{eqnarray}
	Be^{I}&=&-\frac{\sum\limits_{2a \in (Z_2)^{2g}}\exp(-4\pi \mathrm{i} \langle
	\tilde{m}_1^{1},a^{2}\rangle )\Theta'^{2}[n+a](u'+\omega'(\infty^{-}))
	\vartheta^{2}[a](\omega'(\zeta^{+}))}{\sum\limits_{2a \in 
	(Z_2)^{2g}}\exp(-4\pi \mathrm{i} \langle
	m_1^{1},a^{2}\rangle )
	\vartheta^{2}[a](u'+\omega'(\infty^{+}))
	\vartheta^{2}[a](\omega'(\zeta^{+}))}\times\nonumber\\
	&&\sum\limits_{2a \in 
	(Z_2)^{2g}}\exp(-4\pi \mathrm{i} \langle
	m_1^{1},a^{2}\rangle )\frac{\vartheta^{2}[a](u')}{\vartheta^{2}(0)}
	\times\nonumber\\
	&&
	\left(\frac{\vartheta^{2}[a](\omega'(\infty^{+})-\omega'(\zeta^{+}))}{
	\vartheta_{o}^{2}(\omega'(\infty^{+})-\omega'(\zeta^{+}))}
	+\frac{\vartheta^{2}[a](\omega'(\infty^{+})-\omega'(\zeta^{-}))}{
	\vartheta_{o}^{2}(\omega'(\infty^{+})-\omega'(\zeta^{-}))}\right)
	\label{eq17},
\end{eqnarray}
where $n$ follows from $\tilde{m}$ as in Theorem 4.1. The first 
fraction in (\ref{eq17}) is again the quotient of polynomials of 
degree $g-1$ in $\zeta$ for the same reasons as above. But since the 
quotient must vanish for $\zeta\to\infty$, the leading terms in the 
numerator just cancel. It is thus a quotient of polynomials of degree $g-2$ 
or less in the numerator and $g-1$ or less in the denominator.  
To deal with the quotients $
\vartheta^{2}[a](\omega'(\infty^{+})-\omega'(\zeta^{\pm}))/
	\vartheta_{o}^{2}(\omega'(\infty^{+})-\omega'(\zeta^{\pm}))$, we 
define the divisors $T^{\pm}=T_{1}^{\pm}+\ldots+T_{g-1}^{\pm}$ as the 
solutions of the Jacobi inversion problems $\omega'(T^{\pm})-\omega'(Y)=
\omega'(\infty^{+})+\omega'(\zeta^{\pm})$ where $Y$ is the divisor 
$Y=E_{1}+\ldots+E_{g-1}$. Abel's theorem then implies 
for arbitrary $K\in \mbox{C}$
\begin{equation}
	\prod_{i=1}^{g-1}(K-T_{i}^{\pm})(K-\zeta)=
	(K-A^{\pm})^{2}\prod_{i=1}^{g-1}(K-E_{i})-(K-E_{g})\prod_{i=1}^{g}(K-\bar{E}_{i})
	\label{eq17a},
\end{equation}
where 
\begin{equation}
	\zeta-A^{\pm}=\pm \frac{\mu'(\zeta)}{\prod_{i=1}^{g-1}(\zeta-E_{i})}
	\label{eq17b}.
\end{equation}
Let $Q_{j}$ be given by the condition $[Q_{j}+K_{R}]=[a]$, i.e.\ 
$Q_{j}$ is a branch point of $\Sigma'$. Then we get for the quotient 
\begin{equation}
\frac{\vartheta^{2}[a](\omega'(\infty^{+})-\omega'(\zeta^{\pm}))}{
	\vartheta_{o}^{2}(\omega'(\infty^{+})-\omega'(\zeta^{\pm}))}=const 
	\prod_{i=1}^{g-1}\frac{T^{\pm}_{i}-Q_{j}}{T^{\pm}_{i}-E_{1}}
	\label{eq17c},
\end{equation}
where $const$ is a $\zeta$-independent constant. With the help of 
(\ref{eq17a}), it is straight forward to see that 
for $Q_{j}\in Y$, the theta quotient is just 
proportional to $(\zeta-E_{1})/(\zeta-Q_{j})$ whereas for $Q_{j}\notin 
Y$, the term is proportional to 
$(\zeta-E_{1})(Q_{j}-A^{\pm})^{2}/(\zeta-Q_{j})$. Using (\ref{eq17b}) one 
recognizes that the terms containing roots just cancel in 
(\ref{eq17}). The remaining terms are just quotients of polynomials 
in $\zeta$ with maximal degree $g$ in the numerator and $g-2$ in 
denominator. This completes the proof.

\textbf{Remark:} The remaining dependence on $G$ through $u'$ and 
$I$ can only be 
eliminated by differentiating relation (\ref{sing7}) $g$ times. If we 
prescribe e.g.\ the function $b$ on a given Riemann surface (this 
just reflects the fact that the function  $G$ can be freely chosen in 
(\ref{sing7})), we can read off $e^{2U}$ from (\ref{eq13}). To fix the  
constants related to $G$ in (\ref{eq13}) one needs to know the Ernst 
potential and $g-1$ derivatives at some point on the 
axis where the Ernst potential is regular, e.g.\ at the origin or at 
infinity, where one has to prescribe the multipole moments.
If the Ernst potential were known on some regular part 
of the axis, one could use (\ref{eq13}) to read off the Riemann 
surface (genus and branch points).  Equation (\ref{sing7}) is then an 
integral equation for $G$ for known sources. This just reflects a 
result of \cite{hauser} that the Ernst potential for known sources can 
be constructed via Riemann-Hilbert techniques if it is known on some 
regular part of the axis.

In practice it is difficult to express the coefficients in the 
polynomials $P_{i}$ via the constants $u_{i}'$ and $I$, and it will be 
difficult to get explicit expressions. We will therefore concentrate on 
the general structure of the relation (\ref{eq13}), its implications 
on the multipoles and some instructive examples. 
Let us first consider the case genus 1 which is not generically 
equatorially symmetric. In this case the Riemann 
surface $\Sigma'$ is of genus 0. One can use formula (\ref{sing7}) for 
the axis potential if one replaces the theta functions by 1. We thus 
end up  with
\begin{equation}
f\bar{f} -2b \frac{\zeta-\alpha_{1}}{\beta_{1}}
e^{I}=e^{2I}
	\label{bound4}.
\end{equation}
Here the only remaining $G$-dependence is in $I$. If $f_{0}=f(0,0)$  is 
given, $e^{I}$ follows from $f_{0}\bar{f}_{0}+2b_{0}
\alpha_{1}e^{I}/\beta_{1}=e^{2I}$,  if
the in general non-real mass $M$ is known, the constant $e^{I}$ follows 
from $1+4\mbox{Im}M e^{I}/\beta_{1}=e^{2I}$. 
In the latter case the imaginary part of the 
Arnowitt-Deser-Misner mass (this corresponds to a NUT-parameter)
will be sufficient. Differentiating (\ref{bound4}) 
once will lead to a differential 
relation between the real and the imaginary part of the Ernst 
potential which holds for all $G$, which means it reflects only the 
impact of the underlying Riemann surface on the structure of the 
solution. 

\textbf{Remark:} For equatorially symmetric solutions, 
one has on the positive axis the relation $f(-\zeta)\bar{f}(\zeta)=1$ 
(see \cite{panos,meinel}, this is to be understood in the following way: 
the function $|\zeta|$ is even in $\zeta$, but restricted to positive $\zeta$ it seems 
to be an odd function, and it is exactly this behavior which is 
addressed by the above formula). This leads to the conditions 
\begin{equation}
	P_{1}(-\zeta)=-P_{3}(\zeta), \quad P_{2}(-\zeta)=P_{2}(\zeta)
	\label{eq15.1}.
\end{equation}
The coefficients in the polynomials depend on the $g/2$ integrals 
$\int_{\Gamma}^{}d\tau\ln G \tau^{2i}/\mu'(\tau)$ ($i=0,\ldots, g/2-1$
and the branch 
points. 

The simplest interesting example is genus 2, where we get 
with $E_{1}^{2}=\alpha+\mathrm{i}\beta$
\begin{equation}
	f\bar{f}(\zeta-C_{1})+\frac{\sqrt{2}}{C_{2}}(\zeta^{2}-
	\alpha-C_{2}^{2})b=\zeta+C_{1}
	\label{eq16.1},
\end{equation}
i.e.\ a relation which contains two real constants $C_{1}$, $C_{2}$ 
related to $G$. In 
case that the Ernst potential at the origin is known, one can express 
these constants via $f_{0}$. A relation of this type, which is as 
shown typical for the whole class of solutions, was observed in the  
first paper of 
\cite{neugebauermeinel1} for the rigidly rotating dust disk. 

\section{Differential relations in the whole spacetime}\label{sec.5}
The considerations on the axis have shown that it is possible there to 
obtain relations between the real and the imaginary part of the Ernst 
potential which are independent of the function $G$ and thus reflect 
only properties of the underlying Riemann surface. The found 
algebraic relations contain however $g$ real constants related to the function 
$G$, which means that one has to differentiate $g$ times to get a 
differential relation which is completely free of the function $G$. 
These constants were just the integrals $u'$ and $I$ which are only 
constant with respect to the physical coordinates on the axis where 
the Riemann surface $\Sigma$ degenerates. Thus one cannot hope to get 
an algebraic relation in the whole spacetime as on the axis. Instead 
one has to deal with integral equations or to 
look directly for a differential relation.

To avoid the differentiation of theta functions with respect to a 
branch point of the Riemann surface, we use the algebraic formulation 
of the  hyperelliptic solutions (\ref{eq18}) and (\ref{eq19}). From 
the latter it can also be seen how one could get a relation 
independent of $G$ without differentiation: one can consider the 
equations (\ref{eq18}) and (\ref{eq19}) as integral equations for $G$. 
In principle one could try to eliminate $G$ and $X$ from these 
equations and (\ref{ddd2}). We will not investigate this approach but 
try to establish a differential relation. To this end it 
proves helpful to define the symmetric (in the $K_{n}$) 
functions $S_{i}$ via 
\begin{equation}
	\prod_{i=1}^{g}(K-K_i)=:K^g-S_{g-1}K^{g-1}+...+S_0
	\label{eq22}
\end{equation}
i.e.\ $S_{0}=K_{1}K_{2}\ldots K_{g}$, \ldots, 
$S_{g-1}=K_{1}+\ldots+K_{g}$. The equations (\ref{ddd2}) are bilinear 
in the real and imaginary parts of the $S_{i}$ which are denoted by 
$R_{i}$ and $I_{i}$ respectively.
With this notation we get 

\textbf{ Theorem 6.1:}\\
\emph{ The $x_{i}$ and the Ernst potential $f$ are subject to the 
system of differential equations }
\begin{eqnarray}
	0&=&\left(R_{0}-P_{0}R_{1}+\ldots +P_{0}^{g}(-1)^{g}\right)x_{z}-
	\frac{\mathrm{i}}{2}Q_{2}(P_{0})\nonumber\\
	&&-\frac{\mathrm{i}}{2}(1-x^{2})(\ln f\bar{f})_{z}\left(I_{0}-P_{0}I_{1}+\ldots
	+(-1)^{g-1}I_{g-1}P_{0}^{g-1}\right)
	\label{eq23}
\end{eqnarray}
\emph{and for $g>1$} 
\begin{eqnarray}
	x_{j,z} & = &x_{z} \left((-1)^{j+1}R_{j+1}+\ldots+P_{0}^{g-j-1}\right)
	 -\mathrm{i}(x_{j+1}+\ldots +xP_{0}^{g-j-2})\nonumber  \\
	 &  &-\frac{\mathrm{i}}{2}(1-x^{2})(\ln 
	 f\bar{f})_{z}(
	 ((-1)^{j+1}I_{j+1}+\ldots-P_{0}^{g-j-2}I_{g-1})
	\label{ddd35}.
\end{eqnarray}

\textbf{Proof:}\\
Differentiating (\ref{ddd2}) with 
respect to $z$ and eliminating the derivatives of the $K_{i,z}$ via 
the Picard-Fuchs relations (\ref{ddd19}), 
we end up with a linear system of equations for the 
derivatives of the $x_{i}$ and $x$ which can be solved in standard 
manner. The Vandemonde-type determinants can be expressed via the 
symmetric functions. For $x_{z}$ one gets (\ref{eq23}). The equations 
for the $x_{j,z}$ are bilinear in the symmetric functions. They can be 
combined with (\ref{eq23}) to (\ref{ddd35}).

\textbf{Remark:} If one can solve (\ref{ddd2}) for the 
$K_{i}$, the equations (\ref{eq23}) and (\ref{ddd35}) will be  a non-linear 
differential system in $z$ (and 
$\bar{z}$ which follows  from the reality properties) for the 
$x_{i}$, $x$  and $f$ which only contains the branch points of the Riemann 
surface as parameters. 

For the metric function $a$, we get with (\ref{eq24})

\textbf{ Theorem 6.2:}\\
\emph{ The metric function $a$ is related to the functions $x_{i}$ and 
$S_{i}$ via 
\begin{equation}
	Z=\frac{\mathrm{i}x_{g-2}}{1-x^{2}}-I_{g-1}-\frac{\mathrm{i}x\zeta}{1-x^{2}}
	\label{eq26}.
\end{equation}
for $g>1$ and 
\begin{equation}
		Z=-I_{0}+\frac{\mathrm{i}x(\alpha_{1}-\zeta)}{1-x^{2}}
	\label{eq26a}
\end{equation}
for $g=1$.}

\textbf{Proof:}\\
To express the function $Z$ via the divisor $X$, we define  the divisor 
$T=T_{1}+\ldots+T_{g}$ as the solution of the Jacobi inversion problem 
$\omega(T)=\omega(X)+\omega(P)$ where $P$ is in the vicinity of 
$\infty^{-}$ (only terms of first order in the local parameter near $\infty^{-}$ 
are needed). Using the 
formula for root functions (\ref{root2}), we get for the quantity $Z$ 
in (\ref{eq24})
\begin{equation}
	Z=\frac{\mathrm{i}}{2}D_{\infty^{-}}\ln\prod_{i=1}^{g}\frac{T_{i}-
	\bar{P}_{0}}{T_{i}-P_{0}}
	\label{eq25}.
\end{equation}
Applying Abel's theorem to the definition of $T$ and expanding in 
the local parameter near $\infty^{-}$, we end up with (\ref{eq26}) for 
general $g>1$ and with (\ref{eq26a}) for $g=1$.

\textbf{Remark:} \\
1. For $g>1$ 
equation (\ref{eq26})  can be used to replace the relation for 
$x_{g-2,z}$ in 
(\ref{ddd35}) since the latter is identically fulfilled with 
(\ref{eq26}) and (\ref{vac9}).

2. An interesting limiting case is $G\approx 1$ where 
$f\approx 1$, i.e.\ the limit where the solution is close to Minkowski 
spacetime. By the definition (\ref{eq18}), the divisor $X$ is in this 
case approximately equal to $D$. Thus the symmetric functions in 
(\ref{ddd35}) and (\ref{eq23})  can be considered as 
being constant and given by the branch points $E_{i}$.  Relation 
(\ref{eq26}) implies that the quantity $Z$ is approximately equal to 
$I_{g-2}$ in this limit, i.e.\ it is mainly equal to the constant $a_{0}$ 
in lowest order. 
Since the differential system (\ref{ddd35}) and (\ref{eq23}) is linear 
in this limit, it is straight forward to establish two real differential 
equations of order $g$ for the real and the imaginary part of the 
Ernst potential. In principle this works also in the non-linear case, 
where sign ambiguities in the solution of (\ref{eq18}) can be fixed by
the Minkowskian limit.

To illustrate the above equations we will first consider the elliptic 
case. This is the only case where one can establish an algebraic 
relation between $Z$ and $b$ independent of $G$. 
Equations (\ref{ddd2}) lead  to 
\begin{eqnarray}
	(1-x^{2})R_{0} & = & \alpha_{1}-\zeta x^{2}
	\nonumber  \\
	(1-x^{2})S_{0}\bar{S}_{0} & = & E_{1}\bar{E}_{1}-P_{0}\bar{P}_{0}x^{2}
	\label{eq27}.
\end{eqnarray}
Formula (\ref{eq26a}) takes with 
(\ref{eq27}) (the sign of $I_{0}$ is fixed by the condition that 
$I_{0}=-\beta_{1}$ for $x=0$) the form 
\begin{equation}
	(1-x^{2})Z=\mathrm{ i}x(\alpha_{1}-\zeta)+\sqrt{(1-x^{2})(\beta_{1}^{2}-
	\rho^{2}x^{2}) -x^{2}(\alpha_{1}-\zeta)^{2}}
	\label{eq28}.
\end{equation}
This relation holds in the whole spacetime for all elliptic 
potentials, i.e.\ for all possible choices of $G$ in (\ref{rel1}). 
This implies that one can only solve boundary value problems on 
elliptic surfaces where the boundary data at some given contour $\Gamma_{z}$ 
satisfy condition (\ref{eq28}). 

In the case genus 2, we get for (\ref{ddd2})
\begin{eqnarray}
	(1-x^2)R_{1} & = & \alpha_1+\alpha_2-\zeta x^2 +xx_0,
	\nonumber \\
	(1-x^2)(R_{1}^{2}+I_{1}^{2}+2R_{0}) & = & (\alpha_1+\alpha_2)^2+2\alpha_1\alpha_2
	+\beta_1^2+\beta_2^2-
	x_0^2-x^2(\rho^2+\zeta^2)+4\zeta xx_0,
	\nonumber \\
	(1-x^2)(R_{1}R_{0}+I_{1}I_{0}) & = & \alpha_1\alpha_2(\alpha_1+\alpha_2)+\alpha_1\beta_2^2
	+\alpha_2\beta_1^2-\zeta x_0^2+(\rho^2+\zeta^2)xx_0,
	\nonumber \\
	(1-x^2)(R_{0}^{2}+I_{0}^{2}) & = & (\alpha_1^2+\beta_1^2)(\alpha_2^2+\beta_2^2)
	-(\rho^2+\zeta^2)x_0^2
	\label{bound33}.
\end{eqnarray}
The aim is to determine the $S_{i}$ and $x_{0}$ from (\ref{bound33}) 
and 
\begin{equation}
	(1-x^{2})(Z+I_{1})=\mathrm{i}x_{0}-\zeta \mathrm{i}x,
	\label{bound33.1}
\end{equation}
and to eliminate these 
quantities in 
\begin{equation}
	(R_{0}-P_{0}R_{1}+P_{0}^{2})x_{z}=\frac{\mathrm{i}}{2}(x_{0}+P_{0}x) 
	+\frac{\mathrm{i}}{2}(1-x^{2})(\ln f\bar{f})_{z}(I_{0}-P_{0}I_{1})
	\label{eq29}
\end{equation}
which follows from (\ref{eq23}).

\textbf{Remark:} Boundary value problems\\
Since the above 
relations will hold in the whole spacetime, it is possible to 
extend them to an arbitrary smooth boundary $\Gamma_{z}$, where the 
Ernst potential may be singular (a jump discontinuity) and 
where one wants to prescribe 
boundary data (combinations of $f$, $f_{z}$). 
If these data are of sufficient differentiability (at 
least $C^{g,\alpha}(\Gamma_{z})$), we can check the 
solvability of the problem on a given surface with the above formulas. 
The conditions on the differentiability of the boundary 
data can be relaxed by working directly with the equations 
(\ref{eq18}) and (\ref{eq19}) which can be considered as integral 
equations for $\ln G$.  The latter is not very 
convenient if one wants to construct explicit solutions, but it makes 
it possible to treat boundary value problems where the boundary data 
are H\"older continuous.
We will only work with the differential relations and 
consider merely the derivatives tangential to $\Gamma_{z}$ in (\ref{ddd35}) 
to establish the desired differential relations between $a$, $b$ and $U$. 
One ends up with two differential equations which involve only $U$, 
$b$ and derivatives. The aim is to construct the spacetime which 
corresponds to the prescribed boundary data from these relations. To 
this end one has to integrate the differential relations using the 
boundary conditions.
Integrating one of these equations, one gets 
$g$ real integration constants which cannot  be 
freely chosen since they arise from 
applying the tangential derivatives in (\ref{ddd35}). Thus they have 
to be fixed in a way that the integrals on the right-hand side of (\ref{eq18}) 
are in fact the $b$-periods of the second integral on the right-hand 
side of (\ref{eq18}) and that (\ref{eq19}) holds.
The second 
differential equation arises from the use of normal 
derivatives of the Ernst potential in (\ref{eq23}). To satisfy the 
$b$-period condition (\ref{eq18}), one has to fix a free 
function in the integrated form of the corresponding differential equation. 
Thus one has to complement the two differential equations following 
from (\ref{eq23}) with an integral equation which is obtained by 
eliminating $G$ from e.g.\ $\tilde{u}_{1}$ and $\tilde{u}_{2}$ in 
(\ref{eq18}). 
For given boundary data, the system following from 
(\ref{eq18}) may in principle be integrated to give $e^{2U}$ and $b$ in 
dependence of the boundary data. Then the in general non-linear
integral equation will establish whether the boundary data are 
compatible with the considered Riemann surface. This is typically a rather 
tedious procedure. There is however a class of problems where it is 
unnecessary to use this integral equation. In case that the 
differential equations hold for an arbitrary function $e^{2U}$, the 
integral equation will only be used to determine this 
metric function, but the boundary value problem will be always 
solvable (locally). This offers a constructive approach to solve boundary 
value problems without having to consider non-linear integral equations.

\section{Counter-rotating disks of genus 2}\label{sec.6}
Since it is not very instructive to establish the differential 
relations for genus 2 in the 
general case, we will concentrate in this section on the form 
these equations take in the equatorially symmetric case for 
counter-rotating dust disks.
In this case, the solutions are parametrized by 
$E_{1}^{2}=\alpha+\mathrm{i}\beta$. 
We will always assume in the following that the boundary data are 
at least $C^{2}(\Gamma_{z})$ (the normal derivatives of the
metric functions have a jump at the 
disk, but the tangential derivatives are supposed to exist up to at 
least second order).
Putting $s=be^{-2U}$ and $y=e^{2U}$, we get for (\ref{eq29}) for 
$\zeta=0$, $\rho\leq 1$
\begin{eqnarray}
	\mathrm{ i}x_{0} & = & \left(R_{0}-\rho^{2}-sI_{0}\right)\frac{b_{\zeta}}{y}-
	\rho\left(R_{1}-sI_{1}\right)\frac{b_{\rho}}{y}-\left(s(R_{0}-\rho^{2})
	+I_{0}\right)
	\frac{y_{\zeta}}{y}+\rho\left(sR_{1}+I_{1}\right)\frac{y_{\rho}}{y},
	\nonumber \\
	\rho s & = & \left(R_{0}-\rho^{2}-sI_{0}\right)\frac{b_{\rho}}{y}
	+\rho\left(R_{1}-sI_{1}\right)\frac{b_{\zeta}}{y}-
	\left(s(R_{0}-\rho^{2})+I_{0}\right)
	\frac{y_{\rho}}{y}-\rho\left(sR_{1}+I_{1}\right)\frac{y_{\zeta}}{y}
	\label{achse66},\quad\quad
\end{eqnarray}
where the $S_{i}$ and $\mathrm{i}x_{0}$ are taken from (\ref{bound33}) and 
(\ref{bound33.1}). Since counter-rotating dust disks are subject to the boundary 
conditions (\ref{eq9}), we can replace the normal derivatives in 
(\ref{achse66}) via (\ref{eq9}) which leads to a differential system 
where only tangential derivatives at the disk occur. With 
(\ref{bound33}) and (\ref{bound33.1}) we get 
\begin{eqnarray}
	\mathrm{ i}x_{0}+(Z-\mathrm{i}x_{0})\frac{R_{0}-\rho^{2}}{I_{1}R} & = & 
	\left(\rho-\frac{R^2+\rho^2+\delta y^{2}}{2R\rho}
	\frac{R_{0}-\rho^{2}}{I_{1}}\right)\left((-Z+
	\mathrm{i}x_{0})\frac{y_{\rho}}{y}-
	sZ\frac{b_{\rho}}{y}\right)
	\label{achse69}, \\
	\rho s\left(1-\frac{Z}{R}\right) & = & 
	\left(\frac{R_{0}-\rho^{2}}{I_{1}}-\frac{R^2+\rho^2+\delta y^{2}}{2R}
	\right)\left(sZ\frac{y_{\rho}}{y}+(-Z+\mathrm{i}x_{0})\frac{b_{\rho}}{y}
	\right).\quad\quad
	\label{achse70}
\end{eqnarray}
With $I_{1}=\mathrm{i}x_{0}/(1-x^{2})-Z$ and
\begin{equation}
	R_{0}=\frac{\mathrm{i}x_{0}Z-\alpha-\frac{\rho^{2}}{2}}{1-x^{2}}-
	\frac{Z^{2}-\rho^{2}}{2},
	\label{eq40}
\end{equation}
the function $\mathrm{i}x_{0}$ follows from
\begin{equation}
	R_{0}^{2}+\frac{(R_{0}-\rho^{2})^{2}}{I_{1}^{2}}
	\frac{x^{2}x_{0}^{2}}{(1-x^{2})^{2}}=
	\frac{\alpha^{2}+\beta^{2}-\rho^{2}x_{0}^{2}}{1-x^{2}}
	\label{eq41},
\end{equation}
i.e.\ an algebraic equation of fourth order for $\mathrm{i}x_{0}$ which 
can be uniquely solved by respecting the Minkowskian limit. Thus 
(\ref{achse69}) and (\ref{achse70}) are in fact a differential system 
which determines $b$ and $y$ in dependence of the angular velocity 
$\Omega$.

\subsection{Newtonian limit}\label{subsec.6.1}
For illustration we will first study the Newtonian limit of the 
equations (\ref{achse70}) (where counter-rotation does not play a 
role). 
This means  we are looking for dust disks with an 
angular velocity of the form $\Omega=\omega q(\rho)$ where 
$|q(\rho)|\leq 1$ for $\rho\leq 1$, and where the dimensionless 
constant $\omega<<1$. Since we have put the radius $\rho_{0}$ of the 
disk equal to 1, $\omega=\omega \rho_{0} $ is the upper limit for the 
velocity in the disk. The 
condition $\omega<<1$ just means that the maximal velocity in the disk 
is much smaller than the velocity of light which is equal to 1 in the 
used units. An expansion in $\omega$ is thus equivalent to a standard 
post-Newtonian expansion.  Of course there may be dust disks of 
genus 2 which do not have such a limit, but we will study in the 
following which constraints are imposed by the Riemann surface on the 
Newtonian limit of the disks where such a limit exists.

The invariance of the metric (\ref{vac1}) under the transformation 
$t\to-t$ and $\Omega \to -\Omega$ implies that $U$ is an even function 
in $\omega$ whereas $b$ has to be odd. Since we have chosen an 
asymptotically non-rotating frame, we can make the ansatz 
$y=1+\omega^{2}y_{2}+\ldots$, 
$b=\omega^{3}b_{3}+\ldots$, and $a=\omega^{3}a_{3}+\ldots$. 
The boundary conditions (\ref{eq9}) imply in lowest order 
$y_{2,\rho}=2q^{2}\rho$, the well-known 
Newtonian limit. Since  (\ref{vac10}) reduces to the Laplace equation 
for $y_{2}$ in order $\omega^{2}$, we can use the methods of 
section (\ref{sec.2}) to construct the corresponding solution. 
In order $\omega^{3}$, the boundary conditions (\ref{eq9}) lead to
\begin{equation}
	b_{3,\rho}=2\rho qy_{2,\zeta}
	\label{diff13a},
\end{equation}
whereas equation (\ref{vac10}) leads to the Laplace equation for 
$b_{3}$. Again we can use the methods of section \ref{sec.2}, but this 
time we have to construct a solution which is odd in $\zeta$ because 
of the equatorial symmetry. In principle one 
can extend this perturbative approach to higher order, where the 
field equations (\ref{vac10}) lead to Poisson equations with 
terms of lower order acting as source terms, and where the boundary 
conditions can also be obtained iteratively from (\ref{eq9}).

With this notation we get \\
\textbf{ Theorem 7.1:}\\
\emph{ Dust disks of genus 2 which have a Newtonian limit, i.e.\ 
a limit in which $\Omega=\omega q(\rho)$ where 
$|q(\rho)|\leq 1$ for $\rho\leq 1$, are either rigidly rotating 
($q=1$) or $q$ 
is a solution to the integro-differential equation
\begin{equation}
	b_{3}=\left((R_{0}^{0}-\rho^{2})2q-\kappa\right)y_{2,\zeta}
	\label{eq30}
\end{equation}
where in the first case $I_{1}^{0}/R_{0}^{0}=2\omega$ and in the 
second $I_{1}=\kappa \omega$ with $R_{0}^{0}$ and $\kappa$ being 
$\omega$ independent constants.}

\textbf{Proof:}\\
Since the right hand side of (\ref{eq18}) vanishes, we have 
$K_{i}=E_{i}$ for $\omega\to0$, and thus $a_{0}=I_{1}$ up to at least 
order $\omega^{3}$. Keeping only terms in lowest order and denoting 
the corresponding terms of the symmetric functions by $S_{i}^{0}$, 
we obtain for (\ref{achse70})
\begin{equation}
	 \omega^{3}b_{3}=y_{2,\zeta}
	\left(2q(R^{0}_{0}-\rho^{2})\omega^{3}
	-\omega^{2}I^{0}_{1}
	 \right)
	\label{diff17}.
\end{equation}
The second equation (\ref{achse70}) 
involves $b_{3,\zeta}$ and is thus of higher order.
If  (\ref{diff17}) holds, this equation will be automatically fulfilled.

The $\omega$-dependence in (\ref{diff17}) implies that 
$R_{0}^{0}$, $I_{1}^{0}$ and thus the branch points must depend 
on $\omega$. Since $y_{2,\zeta}$ is proportional to the density in 
the Newtonian case, it must not vanish identically. The possible 
cases following from  equation (\ref{diff17}) are constant $\Omega$ 
or (\ref{eq30}). Using 
(\ref{eq2}) and (\ref{eq4}), one can express $U_{\zeta}$ directly 
via $\Omega$ which leads to 
\begin{equation}
	y_{2,\zeta}=\frac{4}{\pi}\int_{0}^{1}\frac{\mathrm{d}\rho'}{\rho+\rho'}
	\partial_{\rho'}(q^{2}\rho'{}^{2}) K(k)
\end{equation}
with $k=2\sqrt{\rho\rho'}/(\rho+\rho')$.
Thus (\ref{eq30}) is in fact an integro-differential equation for 
$q$. This completes the proof.

\subsection{Explicit solution for constant angular velocity and 
constant relative density}\label{subsec.6.2}
The simplifications of the Newtonian equation (\ref{diff17})  for 
constant $\Omega$ give rise to the hope that a generalization of rigid 
rotation to the relativistic case might be possible which we will 
check in the following. Constant $\gamma/\Omega$ makes it in fact
possible to avoid the solution of a differential equation and leads 
thus to the simplest example. We restrict ourselves to the case of 
constant relative density,
$\gamma=const$.
The structure of equation (\ref{achse70}) suggests 
that it is sensible to choose the constant $a_{0}$ as 
$a_{0}=-\gamma/\Omega$ since in this case $Z=R$. This is the only freedom 
in the choice of the parameters $\alpha$ and $\beta$ on the Riemann surface 
one has for $g=2$ since one of the parameters will be fixed as in 
the Newtonian 
case by the condition that the disk has to be regular at its rim. The 
second  parameter will be determined 
as an integration constant of the Picard-Fuchs system.

We get \\
\textbf{ Theorem 7.2:}\\
\emph{ The boundary conditions (\ref{eq9}) and (\ref{eq11}) 
for the counter-rotating dust disk with 
constant $\Omega$ and constant $\gamma$
are satisfied by an Ernst potential of 
the form (\ref{rel1}) on a hyperelliptic Riemann surface of genus 2 
with the branch points specified by 
\begin{equation}
	\alpha=-1+\frac{\delta}{2}, \quad 
	\beta=\sqrt{\frac{1}{\lambda^2}+\delta-\frac{\delta^2}{4}}
	\label{eq37a}.
\end{equation}
The parameter $\delta$ varies between $\delta=0$ (only one component) 
and $\delta=\delta_{s}$, 
\begin{equation}
	\delta_{s}=2\left(1+\sqrt{1+\frac{1}{\lambda^{2}}}\right)
	\label{eq37b},
\end{equation}
the static limit.
The function $G$ is given by 
\begin{equation}
	G(\tau)=\frac{\sqrt{(\tau^{2}-\alpha)^2+\beta^2}+\tau^{2}+1}{
	 \sqrt{(\tau^{2}-\alpha)^2+\beta^2}-(\tau^{2}+1)}
	\label{eq38}.
\end{equation}}
This is the result which was announced in  \cite{prl2}.

\textbf{Proof:}\\
The proof of the theorem is performed in several steps.\\
1. Since the second factor on the right-hand side in (\ref{achse70}) 
must not vanish in the Newtonian limit, we find that for $Z=R$
\begin{equation}
	\frac{R_{0}-\rho^{2}}{I_{1}}=\frac{Z^2+\rho^2+\delta y^{2}}{2Z}
	\label{eq31}.
\end{equation}
With this relation it is possible to solve (\ref{eq40}) and 
(\ref{bound33.1}),
\begin{eqnarray}
	\mathrm{i}x_{0}&=&\frac{Z(\rho^2+2 
	\alpha -\delta y^2 (1-x^{2}))}{Z^2-\rho^2-\delta y^2}\label{eq32}
	\\
		\frac{\delta^2 y^2}{2}(1-x^{2}) & = & -\frac{1}{\lambda}\left(\frac{1}{\lambda}
	-\delta y\right) +\delta\left(\alpha+\frac{\rho^2}{2}\right)+
	\frac{\frac{1}{\lambda}-\delta y}{\sqrt{\frac{1}{\lambda^2}+\delta \rho^2}} 
	\sqrt{\left(\frac{1}{\lambda^2}
	-\alpha\delta\right)^2+\delta^2 \beta^2}.
	\nonumber
\end{eqnarray}
One may easily check that equation (\ref{achse69}) is identically 
fulfilled with these settings. Thus the two differential equations 
(\ref{achse69}) and (\ref{achse70}) are satisfied for an unspecified 
$y$ which implies that the boundary value problem for the rigidly 
rotating dust disk can be solved on a Riemann surface of genus 2 (the 
remaining integral equation which we will discuss below determines then $y$). 

2. To establish  the integral equations which determine 
the function $G$ and the metric potential $e^{2U}$, we use equations 
(\ref{eq18}).  Since we have 
expressed above the $K_{i}$ as a function of $e^{2U}$ alone, the 
left-hand sides of (\ref{eq18}) are known in dependence of $e^{2U}$. It 
proves helpful to make explicit use of the 
equatorial symmetry at the disk. By construction the Riemann surface 
$\Sigma$ is for $\zeta=0$ invariant under the involution $K\to -K$. 
This implies that the theta functions factorize and can be expressed 
via theta functions on the covered surface $\Sigma_{1}$ given by 
$\mu_{1}^{2}(\tau)=\tau(\tau+\rho^{2})((\tau-\alpha)^{2}+\beta^{2})$ 
and the Prym variety $\Sigma_{2}$ (which is here also a Riemann surface) 
given by 
$\mu_{2}^{2}(\tau)=(\tau+\rho^{2})((\tau-\alpha)^{2}+\beta^{2})$ (see 
\cite{algebro,prd2} for details). On these surfaces we 
define the divisors $V$ and $W$ respectively via 
\begin{equation}
	u_v=\frac{1}{\mathrm{i}\pi}\int_{0}^{-\rho^{2}}\frac{\ln 
	G(\sqrt{\tau}) d\tau}{\mu_{1}(\tau)}
	=:\int_{0}^{V}\frac{d \tau}{\mu_1}, \quad u_w= \frac{1}{\mathrm{i}\pi}
	\int_{-\rho^{2}}^{-1}\frac{\ln 
	G(\sqrt{\tau}) d\tau}{\mu_{2}(\tau)}
	=:\int_{\infty}^{W}\frac{d \tau}{\mu_2}
	\label{achse94}.
\end{equation}
For the Ernst potential we get 
\begin{equation}
	\ln f\bar{f}=-\ln \left(1-\frac{2\mathrm{i}x_0}{Z(1-x^2)}\right)+
	\int_{0}^{V}\frac{\tau d\tau}{\mu_1}-I_v
	\label{achse99a}.
\end{equation} 
where $I_{v}=\frac{1}{2\pi\mathrm{i}}\int_{0}^{-\rho^{2}}\frac{\ln 
	G(\sqrt{\tau})\tau 
\mathrm{d}\tau}{\mu_{1}(\tau)}$.

3. Using Abel's theorem and (\ref{eq18}), we can express $V$ and $W$ 
by the divisor  $X$ which leads to 
\begin{equation}
	V=-\frac{\rho^2 x_0^2}{Z^2(1-x^2)-2Z\mathrm{i}x_0}
	\label{eq33}
\end{equation}
and 
\begin{equation}
	W+\rho^2=-\frac{1}{x^2}\left(
	Z^2(1-x^2)-2Z\mathrm{i}x_0 -x_0^2\right)
	\label{eq34}.
\end{equation}
 
4. Since $V$ and $I_{v}$ vanish for $\rho=0$, we can use 
(\ref{achse99a}) for $\rho=0$ to determine the integration constant of the
Picard-Fuchs system. We get with (\ref{eq32})
\begin{equation}
	\beta^2=\frac{1}{\lambda^2}-\delta\alpha+\frac{\delta^2}{4}
	\label{eq35}.
\end{equation}

5.  Since  $V$ in (\ref{eq33}) is  with (\ref{eq32}) a rational function of 
$\rho$, $\alpha$ and $\beta$ and does not depend on the metric 
function $e^{2U}$, we can use the first equation in (\ref{achse94}) to 
determine $G$ as the solution of  an Abelian integral which is 
obviously linear. With $G$ determined in this way, the second equation 
in (\ref{achse94}) can then be used to calculate $e^{2U}$ at the disk 
which leads to elliptic theta functions (see also \cite{prd2}). (In 
the general case, one would have to eliminate $e^{2U}$ in the 
relations for $u_{v}$ and $u_{w}$ to end up with a non-linear integral 
equation for $G$.)

The integral equation following from (\ref{achse94}),
\begin{equation}
	\int_{0}^{V}\frac{\mathrm{d}\tau}{\mu_{1}(\tau)}=\frac{1}{\mathrm{i}\pi}
	\int_{0}^{-\rho^{2}}\frac{\ln G}{\mu_{1}(\tau)}\mathrm{ d}\tau
	\label{eq36}
\end{equation}
is an Abelian equation
and can be solved in standard manner by integrating both sides of 
the equation with a factor $1/\sqrt{K-r}$ from $0$ to $r$ where 
$r=-\rho^{2}$. With (\ref{eq33}) we get for what is essentially an 
integral over a rational function
\begin{equation}
	 G(K)=\frac{\sqrt{(K-\alpha)^2+\beta^2}+K-\alpha+\frac{\delta}{2}}{
	 \sqrt{(K-\alpha)^2+\beta^2}-(K-\alpha+\frac{\delta}{2})}
	\label{achse114}.
\end{equation}

6. The condition $G(-1)=1$ excludes ring singularities at the rim of the 
disk and leads to a continuous potential and density there. It 
determines the last degree of freedom in (\ref{achse114}) to
\begin{equation}
	\alpha=-1+\frac{\delta}{2}
	\label{eq37}.
\end{equation}

7. The static limit of the counter-rotating disks is reached for 
$\beta=0$, i.e.\ the value $\delta_{s}$.
This completes the proof.

\textbf{Remark:} \\
1. It is interesting to note that there are algebraic 
relations between  $a$, $b$ and 
$e^{2U}$ though they are expressed via theta functions, i.e.\ transcendental 
functions, also at the disk.

2. It is an interesting question whether there exist disks with 
non-constant $\gamma/\Omega$  or $\delta$ for genus 2 in the vicinity of the 
above class of solutions. Whereas this is rather straight forward for 
a non-constant $\delta$ if $\gamma/\Omega$ are constant, it is less 
obvious if the latter does not hold. 
This means that one looks for given $\delta$
for solutions with 
\begin{equation}
	\frac{\gamma}{\Omega}=C_{0}+\epsilon p(\rho)
	\label{diff1}
\end{equation}
where $C_0$ is a constant,  $|p|\leq 1$ is a  function of 
$\rho$,  and where
$\epsilon<<1$ is a small dimensionless parameter. We can assume that $p$ 
is not identically constant since this would only lead to a 
reparametrisation of the above solution. To check if there are 
solutions for small enough $\epsilon$, one has to redo the steps 
in the proof of theorem 7.2 in first order of $\epsilon$ by expanding 
all quantities in the form $y=\bar{y}+\epsilon \hat{y}+...$. Doing 
this one recognizes that equation (\ref{achse69}) becomes 
a linear first order differential equation for $p$ of the form 
$p_{\rho}+F(\rho)p=0$ where $F$ is given by the solution for rigid 
rotation. For a solution $p$ to this equation, the remaining steps can 
be performed as above. It seems possible to use the theorem on 
implicit functions to prove the existence of solutions for genus 2 in 
the vicinity of constant $\gamma/\Omega$, but this is beyond the 
scope of this article.

\subsection{Global regularity}\label{subsec.regularity}
In Theorem 7.2 it was shown that one can identify an Ernst potential 
on a genus 2 surface which takes the required boundary data at the disk. 
One has to notice however 
that this is only a local statement which does not ensure one has 
found the desired global solution which has to be regular in the whole 
spacetime except at the disk. It was shown in \cite{prl,prd2} that this is 
the case if $\Theta(\omega(\infty^{-})+u)\neq 0$. In the Newtonian 
theory (see section (\ref{sec.2})), the boundary value problem could be 
treated at the disk alone because of the regularity properties of the 
Poisson integral. Thus one knows that the above condition will hold in 
the Newtonian limit of the hyperelliptic solutions if the latter exists. 
For physical reasons it is however clear that this will not be the 
case for arbitrary values of the physical parameters: if more and more 
energy is concentrated in a region of spacetime, a black-hole is 
expected to form (see e.g.\ the hoop conjecture \cite{hoop}). The 
black-hole limit will be a stability limit for the above disk 
solutions. Thus one expects that additional singularities will occur 
in the spacetime if one goes beyond the black-hole limit. The task is 
to find the range of the physical parameters, here $\lambda$ and 
$\delta$, where the solution is regular except at the disk.

We can state the \\
\textbf{Theorem 7.3:}\\
\emph{Let $\Sigma'$ be the Riemann surface given by 
$\mu'{}^{2}=(K^{2}-E)(K^{2}-\bar{E})$ and let a prime denote that the 
primed quantity is defined on $\Sigma'$. Let $\lambda_{c}(\delta)$ be 
the smallest positive value $\lambda$ for which $\Theta'(u')=0$. Then 
$\Theta(\omega(\infty^{-})+u)\neq 0$ for all $\rho$, $\zeta$ and 
$0<\lambda <\lambda_{c}(\delta)$ and $0\leq \delta\leq  \delta_{s}$.}\\
This defines the range of the physical parameters where the Ernst 
potential of Theorem 7.2 is regular in the whole spacetime except at 
the disk. Since it was shown in \cite{prl,prd2} 
that $\Theta'(u')=0$ defines the limit in which the solution can be 
interpreted as the extreme Kerr solution, the disk solution is 
regular up to the black-hole limit if this limit is reached.\\
\textbf{Proof:}\\
1. Using the divisor $X$ of (\ref{eq18}) and the vanishing condition 
for the Riemann theta function, we find that 
$\Theta(\omega(\infty^{-})+u)= 0$ is equivalent to the condition that 
$\infty^{+}$ is in 
$X$. The reality of the $\tilde{u}_{i}$ implies that  
$X=\infty^{+}+(-\mathrm{i}z)$. Equation (\ref{eq18}) thus leads to 
\begin{eqnarray} 
\int_{E_1}^{\infty^+}\frac{d\tau}{\mu}+\int_{E_2}^{-\mathrm{i}z}\frac{d\tau}{\mu} 
 -\frac{1}{2\pi \mathrm{i}}\int_{\Gamma}^{}\frac{\ln G d\tau}{\mu}& \equiv & 0
 , \nonumber \\ 
\int_{E_1}^{\infty^+}\frac{\tau d\tau}{\mu}+\int_{E_2}^{-\mathrm{i}z}\frac{\tau 
d\tau}{\mu}  -\frac{1}{2\pi \mathrm{i}}\int_{\Gamma}^{}
\frac{\ln G \tau d\tau}{\mu} & \equiv &0
\label{count55} ,
\end{eqnarray} 
where $\equiv$ denotes equality up to periods. The reality and the 
symmetry with respect to $\zeta$ of the above expressions limits the 
possible choices of the periods. It is straight forward to show that 
$\Theta(\omega(\infty^{-})+u)= 0$ if and 
only if the functions $F_{i}$ defined by 
\begin{eqnarray} 
F_{1}&:=&\int_{E_1}^{\infty^+}\frac{d\tau}{\mu}
+\int_{E_2}^{-\mathrm{i}z}\frac{d\tau}{\mu}
-n_{1}\left(2\oint_{b_{1}}\frac{d\tau}{\mu}+2\oint_{b_{2}}\frac{d\tau}{\mu}
+\oint_{a_{1}}\frac{d\tau}{\mu} +\oint_{a_{2}}\frac{d\tau}{\mu}\right)
-\frac{1}{2\pi \mathrm{i}}\int_{\Gamma}^{}
\frac{\ln G  d\tau}{\mu} ,
\nonumber\\ 
F_{2}&:=&\int_{E_1}^{\infty^+}\frac{\tau d\tau}{\mu}
+\int_{E_2}^{-\mathrm{i}z}\frac{\tau d\tau}{\mu}
-n_{2}\left(2\oint_{b_{1}}\frac{\tau d\tau}{\mu}
+2\oint_{b_{2}}\frac{\tau d\tau}{\mu}
+\oint_{a_{1}}\frac{\tau d\tau}{\mu} +\oint_{a_{2}}\frac{\tau d\tau}{\mu}\right)
\nonumber\\
&&-\frac{1}{2\pi \mathrm{i}}\int_{\Gamma}^{}
\frac{\ln G  \tau d\tau}{\mu}
\label{count55a} 
\end{eqnarray} 
with the cut system of Fig.~1 and with $n_{1,2}\in \mathrm{Z}$ 
vanish for the same values of $\rho$, $\zeta$, $\lambda$, $\delta$. The 
functions $F_{i}$ are both real, $F_{1}$ is even in $\zeta$ whereas 
$F_{2}$ is odd. Thus $F_{2}$ is identically zero in the equatorial 
plane outside the disk.
\\
2. In the Newtonian limit $\lambda \approx 0$, the above expressions 
take in leading order of $\lambda$ the form
\begin{equation}
	F_{1}=\lambda\left((-8n_{1}+1)c_{1}(\rho,\zeta) \ln \lambda-d_{1}(\rho,\zeta) 
	\lambda\right)
	\label{count55b},
\end{equation}
and 
\begin{equation}
	F_{2}=\sqrt{\lambda}\left((-8n_{2}+1)c_{2}(\rho,\zeta) \ln \lambda-
	d_{2}(\rho,\zeta) 
	\lambda^{\frac{3}{2}}\right)
	\label{count55b1},
\end{equation}
where we have used the same approach as in the calculation of the 
axis potential in (\ref{sing7}) (see \cite{prd2} and references given 
therein); the functions $c_{1}$, $d_{1}$ are non-negative whereas 
$c_{2}/d_{2}$ is positive in $\mathrm{C}/\{\zeta=0\}$. Thus the 
$F_{i}$ are zero for $\lambda=0$ which is Minkowski spacetime $f=1$, 
but they are not simultaneously zero for small enough $\lambda$, 
i.e.\ $f$ is regular in the Newtonian regime in 
accordance with the regularity properties of the Poisson integral. 
The $F_{i}$ may vanish however at some value $\lambda_{s}$ for given 
$\rho$, $\zeta$ and $\delta$. Since we are looking for zeros of the 
$F_{i}$ in the vicinity of the Newtonian regime, we may put 
$n_{1,2}=1$ here.\\
3. Let $\mathcal{G}$ be the open domain $C/\{\zeta=0,\rho\leq1 \vee
\rho=0\}$. It is straight forward to check that the $F_{i}$ are a 
solution to the Laplace equation $\Delta F_{i}=0$ with $\Delta = 
4\left(
\partial_{z\bar{z}}+\frac{1}{2(z+\bar{z})}
(\partial_{z}+\partial_{\bar{z}})\right)$ 
for $z,\bar{z}\in\mathcal{G}$. Thus by the maximum principle
the $F_{i}$ do not have an 
extremum in $\mathcal{G}$.\\
4. At the axis for $\zeta>0$, the $\tilde{u}_{i}$ are finite whereas 
the $F_{i}$ diverge proportional to $-\ln \rho$ for all $\lambda$, 
$\delta$. Thus $f$ is always regular at the axis.\\
5. Relation (\ref{eq23}) at the disk can be written in the form 
$(y+A)^2+b^2=B^2$ where $A$ and $B$ are finite real quantities. Thus 
the Ernst potential is always regular at the disk. Due to symmetry 
reasons $F_{2}\equiv \tilde{u}_{2}$ which is non-zero except at the 
rim of the disk. For $F_{1}$ one gets at the disk
\begin{equation}
	F_{1}=\int_{-\rho^{2}}^{\infty^{+}}\frac{d\tau}{\mu_{1}(\tau)}+\int_{0}^{E}
	\frac{d\tau}{\mu_{1}(\tau)}+\int_{0}^{\bar{E}}\frac{d\tau}{\mu_{1}(\tau)} 
	-u_{v}
	\label{v1}.
\end{equation}
With (\ref{eq36}) one can see that $F_{1}$ is always positive at the 
disk.\\
6. Since $F_{1}$ is strictly positive on the axis and the disk and a solution 
to the Laplace equation in $\mathcal{G}$, it is positive in 
$\bar{\mathrm{C}}$ if it is positive at infinity. $F_{1}$ is regular 
for $|z|\to\infty$ and can be expanded as $F_{1}=F_{11}/|z| +o(1/|z|)$ 
where $F_{11}$ can be expressed via quantities on $\Sigma'$. We get 
\begin{equation}
	F_{11}=\frac{1}{2}
	\oint_{b_{1}'}^{}\frac{d\tau}{\mu'}-\frac{1}{2\pi \mathrm{i}} 
	\int_{-\mathrm{i}}^{\mathrm{i}}\frac{\ln Gd\tau}{\mu'}
	\label{count55c}.
\end{equation}
The quantity $F_{11}\equiv 0$ iff $\Theta'(u')=0$. The condition 
$F_{11}>0$ is thus equivalent to the condition that 
$\lambda<\lambda_{c}(\delta)$ where $\lambda_{c}(\delta)$ is the first 
positive zero of $\Theta'(u')$. This completes the proof.

\textbf{Remark:}\\
1. In the second part of the paper we will show that the 
ultrarelativistic limit (vanishing central redshift) 
in the case of a disk with one component is 
given by $\Theta'(u')=0$ for a finite value of $\lambda$. In the 
presence of counter-rotating matter, this limit is however not 
reached, the central redshift diverges for $\lambda=\infty$ and 
$\Theta'(u')\neq0$. This supports the intuitive reasoning that 
counter-rotation makes the solution more static, i.e.\ it behaves more 
like a solution of the Laplace equation with the regularity 
properties of the Poisson integral. \\
2. Since $F_{2}(\rho,0)=0$ for $\rho\geq 1$, the reasoning in 6.~of 
the above proof shows that there will be a zero of 
$\Theta(\omega(\infty^{-})+u)$ and thus a pole of the Ernst potential 
in the equatorial plane for $\lambda>\lambda_{c}(\delta)$ if the theta 
function in the numerator does not vanish at the same point. In the equatorial 
plane the Ernst potential can be expressed via elliptic theta 
functions (see \cite{prd2}) which have first order zeros. Thus 
$F_{11}$ will be negative for $\lambda>\lambda_{c}$ in the vicinity of 
$\lambda_{c}$, and consequently the same holds for $F_{1}$ in the 
equatorial plane at some value $\rho>1$. It will be shown in the 
third article that the spacetime has  a 
singular ring in the equatorial plane in this case. The disk is 
however still regular 
and the imposed boundary conditions are still satisfied. This provides 
a striking example that one cannot treat boundary value problems 
locally at the disk alone in the relativistic case. Instead one has 
to identify the range of the physical parameters where the solution is 
regular except at the disk.

\section{Conclusion}\label{conclusion}
We have shown in this paper how methods from algebraic geometry can be 
successfully applied to construct explicit solutions for boundary 
value problems to the Ernst equation. We have argued that there 
will be differentially rotating dust disks for genus 2 of the 
underlying Riemann surface in addition to the one we could identify 
explicitly. To 
prove existence theorems for solutions to boundary value problems, the 
methods of \cite{up,urs} seem to be better suited since the 
hyperelliptic techniques are limited to finite genus of the Riemann surface. 
Moreover the used techniques at the boundary have to be complemented 
by a proof of global regularity. The finite genus of the Riemann 
surface also restricts the usefulness in the numerical treatment of 
boundary value problems. The methods of \cite{eric} and \cite{lanza} 
are not limited in a similar way and have proven to be highly 
efficient. Thus the real strength of the approach we have presented 
here is the possibility to construct explicit solutions whose physical 
features can then be discussed in analytic dependence on the physical 
parameters up to the ultrarelativistic limit. 
Whether this approach can be generalized to more 
sophisticated matter models or whether the equations can still be 
handled for higher genus will be the subject of further research.

\noindent \emph{ Acknowledgement}\\
I thank J.~Frauendiener, R.~Kerner, D.~Korotkin, H.~Pfister, 
O.~Richter and U.~Schaudt for helpful remarks and 
hints. The work was supported by the DFG and the Marie-Curie program 
of the European Community.

\end{document}